\documentclass[10pt, a4paper]{article}

\usepackage{amssymb}
\usepackage{amsmath}
\usepackage{a4wide}
\usepackage{graphicx}

\usepackage{pstricks}
\usepackage{slashed}
\usepackage{nth}

\newtheorem{Th}{Theorem}[section]
\newtheorem{definition}[Th]{Definition}
\newtheorem{lemma}[Th]{Lemma}
\newtheorem{example}[Th]{Example}

\graphicspath{{images/}}

\title{\textbf{Non-embeddable relational configurations}}

\author{William Edwards \\ \emph{Perimeter Institute, 31 Caroline Street North,} \\ \emph{Waterloo, Ontario, N2L 2Y5, Canada} \\ \texttt{bedwards@perimeterinstitute.ca}}

\date{}

\begin{document}

\maketitle

\begin{abstract}
Relational configurations are defined by the relationships between systems. Individualist configurations are defined by the properties of individual systems. Non-embeddable relational configurations are those for which there is no equivalent individualist description. We suggest that situations in physics where a spacetime description breaks down may be the result of trying to use individualist configurations to model non-embeddable relational configurations. We propose a particular example of a relational configuration space, the 2 instant relation (2IR), whose embeddable configurations describe particle worldlines in a background spacetime. We investigate the properties of these configurations. 
\end{abstract}

\section{Motivation}

We have a strong sense that the objects around us are situated in a 3-dimensional space, yet all that we can directly observe are the objects themselves and their spatial relationships (the distances, angles etc. between them). Is the positioning of the objects in a background space the fundamental reality, with the spatial relationships a derived notion? Or vice versa? This debate is thousands of years old, and its updated version, with spacetime substituted for space, is still open and relevant to modern physics (see e.g. \cite{SmolinBI}). This work focuses on a possibility opened up by the relational view which seems to have been little-explored by physicists. 

Suppose it is the relationships which are fundamental. For definiteness we will imagine that we have a set of $n$ particles, and associated with each pair of particles is a positive real number `distance' $r_{ij}$. In fact, for most such configurations we cannot interpret the $r_{ij}$s as true distances in a  background space: to do so a number of constraints must hold among the $r_{ij}$s. For example, given any three particles, the distances $r_{ij}, r_{jk}, r_{ki}$ must satisfy the triangle inequalities (i.e. $r_{ij}\leq r_{jk} + r_{ki}$ and all cyclic permuations). More generally for all subsets of $3\leq m\leq n$ particles, constraints must be satisfied (both inequalities and equalities), the exact form of which depends on the dimension and geometry of the background space (see \cite{Blumenthal} for details). We will describe these as \emph{embedding constraints} since they determine whether the relational configuration can be embedded into a background. 

If the relationships are fundamental, there is no a priori reason to insist that these constraints hold, and a general \emph{relational configuration} is a specification of $\frac{1}{2}n(n-1)$ real numbers $\{r_{ij}|i,j = 1,\dots,n; i<j\}$. But if the embedding constraints do hold for some reason then the number of degrees of freedom is significantly reduced. For example, if the embedding is into 3-dimensional Euclidean space, there are just $3n-6$ degrees of freedom required to specify the configuration. One way to see this is as follows: once we specify the distances between three reference bodies, then it is enough to specify the distances of each other particle to these three reference bodies in order to determine the distance between any pair of bodies. The relationalist then has an explanation for why our brains give us such a strong sense of objects being situated in a background space: it's computationally easier to keep track of the 3 degrees of freedom associated with each of the bodies in which you are interested, than to keep track of the distance between each pair. 

What if the constraints hold on the majority of the relationships, but not all? The arguments above imply that a description in terms of a background space would still be useful, but in this case such a description would not do justice to all of the relationships. Here are some examples: 

\begin{enumerate}

\item
Suppose the relational configuration $R = \{r_{ij}|i,j = 1\dots n; i<j\}$ does not satisfy the embedding constraints, but would do if $r_{12}$ was replaced with $r'_{12}\gg r_{12}$. We could embed the $n$ particles in a background space in such a way that the distance between each pair matched the value $r_{ij}$ except for the distance between particles 1 and 2, which would be much greater than the true value $r_{12}$. Then, very loosely, we might expect particles 1 and 2 to behave as if they were much closer than they appeared to be in the background space - an apparent example of \emph{non-locality}. 

\item
Suppose that the full relational configuration $R$ does not satisfy the embedding constraints, but that $R' = \{r_{ij}|i,j = 2\dots n; i<j\}$ does ($R'$ is the subset of all pairwise distances between particles 2 through to $n$).  We could embed particles 2 - $n$ in a background space, but there would be no obvious position to localise particle 1 at. Now suppose that the values of $\{r_{1j}|j=2\dots n\}$ evolve so that momentarily the entire configuration $R$ is embeddable. Now we \emph{can} localise particle 1 at a point. Perhaps this could be the basis of an account of \emph{quantum measurement} in a hidden variable theory. 

\item
Suppose that the full relational configuration $R$ evolves so that prior to time $t$ it satisfies the embedding constraints, but after $t$ only $R'\subset R$ (the set of distances which don't involve particle 1) satisfies the constraints. We could embed all $n$ particles in a background up until time $t$, but after this time it would be impossible to usefully localise particle 1 anywhere in the space. Looked at another way, particle 1's worldline would be abruptly terminated, although particle 1 would continue to exist. The premature termination of particle worldlines is one of the defining characteristics of \emph{singularities} in general relativity (see for example \cite{H&E}). 

\end{enumerate}

Of course the examples above provide only hints and intuitions, and they're phrased in terms of relational configurations (the $r_{ij}$s) which would have to give way to something more compatible with relativity. Nonetheless, they suggest the following possibility: the configurations of our universe are fundamentally relational and non-embeddable; for some (perhaps dynamical) reason the configurations of the bodies which we deal with in everyday life are approximately embeddable so that a description in terms of a background space or spacetime is useful; apparent anomalies in, or breakdowns of the spacetime description in QM and GR are artefacts of trying to embed a non-embeddable relational configuration into a background. 

Our goal is to flesh out these intuitions into something more rigorous which could be useful for real physics. This paper describes some tentative first steps in this programme. 
In section 2 we formalise the notions of relational and individualist configurations, which enable us to formulate our aims precisely. 
In section 3 we argue for a particular relativistic generalisation of the relational particle configurations described above which we term the \emph{two-instant relation} or \emph{2IR}. 
In section 4 we investigate general properties of the 2IR. In section 5 we look specifically at the simplest interesting 2IR - one with three systems. 
In section 6 we return to general considerations of relational and individualist configurations, and look at how individualist configurations could be derived from relational configurations, an understanding of which will be crucial in re-deriving existing theories from explicitly relational ones.

\section{Relational and individualist configurations}\label{sec_rel_ind}

\subsection{Basic concepts}\label{sec_RCS_ICS}

In the previous section we described the debate over the nature of space, between \emph{substantivalists} who claim that matter is positioned within an independently existing spatio-temporal framework, and that spatio-temporal relations amongst material bodies are secondary concepts, and \emph{relationalists} who claim that matter and the spatio-temporal relationships amongst different `pieces' of the matter are all that truly exist, and that `space', `spacetime' and similar notions are convenient fictions which enable a more succinct description of these relationships.

The substantivalist view is most naturally implemented in physical theories whose configurations are \emph{individualist}. An individualist configuration consists of an assignment of attributes to the individual systems of the theory. Put another way, the individualist configuration space (ICS) is a Cartesian product of configuration spaces : $\mathcal{I} = \mathcal{I}_1 \times \dots \times \mathcal{I}_n$ where the degrees of freedom of $\mathcal{I}_i$ refer only to attributes of the $i^\text{th}$ system. The configurations of Newtonian gravity, for example, are of this sort: a configuration specifies a vector $\mathbf{x}_i$ to each particle, giving its position in a background space. 
Conversely the relationalist view is most naturally implemented in physical theories whose configurations are \emph{relational}. Two distinct relational configurations differ in assigning different relationships amongst the the individual systems of the theory; attributes are not assigned to these individual systems. The `$r_{ij}$-configurations' described in the previous section, which assign a `distance' to each pair of particles, are relational configutations. No position is assigned to individual particles; it is a meaningless idea for this sort of configuration. 

A special case of individualist configuration arises when we have a space $\mathcal{I}_0$ such that there is a special isomorphism from every $\mathcal{I}_i$ into $\mathcal{I}_0$. This allows us to think of an individualist configuration as assigning every system to a point in the same space $\mathcal{I}_0$. One could argue that what we think of as `space', is the $\mathcal{I}_0$ for an $\mathbf{x}_i$-configuration. More generally, the $\mathcal{I}_i$s may not be isomorphic but there may be some mapping between each pair $\mathcal{I}_i$ and $\mathcal{I}_j$. If such maps do exist then we may be able to derive a relational configuration $R$ from the individualist configuration $I$ and we wind up with a map\footnote{Given an individualist configurations space $\mathcal{I}$ there might be multiple such constructions. For example given the positions of $n$ particles in space we could derive several different relational configurations: instead of distances between particles we might just give angles. } $D:\mathcal{I}\rightarrow \mathcal{R}$ where $\mathcal{R}$ is a relational configuration space (RCS). For example, from a $\mathbf{x}_i$-configuration giving us the position of each particle we can derive a relational $r_{ij}$-configuration consisting of the distance between each pair of particles. 
Typically the map $D:\mathcal{I}\rightarrow \mathcal{R}$ derived from such a construction is not surjective i.e. there exist relational configurations $R\in\mathcal{R}$ which cannot be derived from any individualist configuration $I\in\mathcal{I}$. An example of this appeared in the first section where we saw that many $r_{ij}$-configurations cannot be interpreted as the distances between particles in a background space, i.e. that they can't be derived from $\mathbf{x}_i$-configurations. Those configurations $R \in \mathrm{Image}(D) \subset\mathcal{R}$ will be termed \emph{embeddable} and all other elements of $\mathcal{R}$ will be termed \emph{non-embeddable}. In the cases where all the $\mathcal{I}_i$s are isomorphic to $\mathcal{I}_0$, we might talk about embedding a relational configuration $R$ into the `space' $\mathcal{I}_0$. 

The $r_{ij}$-configurations we have been considering are \emph{explicitly relational configurations}: the configurations are given by directly specifying relationships (in this case `distances') between the individual systems. It is also possible to specify relational configurations in a more roundabout way, beginning with an individualist configuration space $\mathcal{I}$. We then identify a subgroup $G_{\textrm{Rel}}$ of $\textrm{Diff}(\mathcal{I})$ consisting of maps which in some sense `preserve the relational data' contained in the individualist configurations. The relational configurations are then defined to be the orbits of $G_{\textrm{Rel}}$. We will describe these as \emph{individualist-derived relational configurations} or \emph{ID-relational configurations}, in contrast to explicitly relational configurations. Exactly what is meant by `preserving the relational data' will vary from case to case. An example of this sort of construction is the `background independence' or `diffeomorphism invariance' of general relativity: here we begin by giving individualist configurations of metric and matter fields on a background manifold $\mathcal{M}$, but the true physical states are deemed to be the equivalence classes of these individualist configurations under $\mathrm{Diff}(\mathcal{M})$. 
Those who believe that nature is fundamentally relational may find the appearance of ID-relational configurations in fundamental theories to be somewhat unsatisfactory, since they require `auxiliary' and supposedly unphysical individualist configurations for their formulation. For our purposes, ID-relational configurations are unsatisfactory for another reason. The ID-relational construction gives rise to a map $D:\mathcal{I}\rightarrow \mathcal{R}$ which is surjective by the definition of $\mathcal{R}$ (since $\mathcal{R} = \mathcal{I}/G_{\textrm{Rel}}$), and thus there are no non-embeddable states. Since these are the focus of our interest, we will find little use for ID-relational configurations, and will prefer to work with explicitly-relational configurations. 

We next consider a significant difference between individualist and relational configurations which can arise when we consider dynamics. We define dynamics here to mean a map $\phi: \mathcal{C}\times\mathbb{R}\rightarrow\mathcal{C}:: (C_0, t) \rightarrow C_t$ describing time evolution on the configuration space, be it individualist or relational. An ICS is a Cartesian product $\mathcal{I} = \mathcal{I}_1 \times \dots \times \mathcal{I}_n$. Thus one possible dynamics would be obtained by specifying dynamics $\phi_i(t,\mathcal{I}_i)$ individually for each $\mathcal{I}_i$. This is naturally interpreted as a set of isolated systems evolving independently of each other. Typically we are actually interested in theories with interactions, but the dynamics for such a theory can naturally be specified by giving a \emph{default evolution} $\phi_i$ for each individual system, and then describing how interactions perturb each system from this default evolution.  In contrast a relational configuration space is not easily decomposed into parts relating to individual systems, and dynamics is not naturally stated in terms of default evolutions. Dynamics in a theory with relational configurations does not naturally allow for `turning off interactions'. 

\subsection{Our aims formulated precisely}

The intuitions of the introduction can now be stated more precisely. We aim to formulate a theory with the following features: 
\begin{itemize}
\item
The configurations of the theory should be explicitly relational. There should be a subspace of $\mathcal{R}$ whose configurations embed into classical spacetime, but the majority of configurations should not. 
\item
The dynamics of the theory should predict that the large bodies to which classical physics applies remain in configurations which do embed into spacetime, and that these configurations evolve in accordance with classical physics, at least in the regimes where classical physics is well-tested. 
\item
Quantum mechanics should emerge as the description of the interaction between these embeddable bodies and additional non-embeddable bodies. `Measurement', `observers' and similar notions should be emergent concepts and play no role in the theory at a fundamental level. 
\item
Breakdowns of the spacetime description in classical physics such as singularities in GR should be explained as evolutions of configurations of classical bodies from embeddable to non-embeddable. 
\end{itemize}

In this paper we can promise only modest progress on just the first two points. 
Our first task is to specify the appropriate relational configuration space, and we will be guided by the need to re-derive classical physics in appropriate circumstances. This motivates a review of how relational and individualist configurations appear in classical physics. 

\subsection{Relational and individualist configurations in physics}

In the terminology of section \ref{sec_RCS_ICS} we can see that the debate between substantivalists and relationalists is given life by the fact that we directly observe only relational configurations of material bodies (evidence for the relationalist position), but these relational configurations are all embeddable, i.e. derivable from an individualist configuration (evidence for the substantivalist position). One piece of evidence which would settle the debate would be if non-embeddable relational configurations were discovered. We have suggested that the difficulties of maintaining a spacetime description in quantum mechanics and general relativity might constitute exactly such evidence. But we are very far from having established this, and certainly no hints of such evidence were available until the twentieth century. 

In section  \ref{sec_RCS_ICS} we introduced the notion of a default evolution, and noted that such evolution is only naturally accommodated by theories involving individualist configurations. The identification of a default motion is therefore often seen as strong evidence for the primacy of individualist configurations, and the realisation in the sixteenth century that motion could be factored into inertial motion and the perturbing effect of forces naturally resulted in the development of theories employing individualist configurations (Newtonian mechanics) and gave support to the substantivalist conception of space. 

A relationalist response to an apparent default motion was first identified by Mach \cite{Mach}. In loose terms it goes as follows. Suppose the universe is such that we can partition its contents into (i) a small subsystem whose motion we are interested in and (ii) a large number of bodies whose relationships are approximately static over some suitable timescale, which constitute an effective `background' (in Mach's case, these were the `fixed stars'). Then what looks like a default motion of bodies in the small subsystem of interest could be their motion w.r.t. the background bodies. The universe is clearly very large, and our theories employing individualist configurations have only really been tested on small subsystems of the universe. So the possibility is open that our familiar laws employing individualist configurations are approximations which apply only to small subsystems of the universe (the solar system, the galaxy etc.) and that the universe as a whole is best described by unfamiliar laws employing relational configurations. This idea is commonly referred to as `Mach's principle'. 

Mach was specifically addressing the issue of inertial motion, and he envisaged an explicitly relational theory which approximated Newtonian gravitation when describing small subsystems of a large universe of particles. Mach never advanced an explicit implementation of his ideas. This actually had to wait until 1977 and the work of Barbour and Bertotti \cite{BB77}. Their theory (henceforth BB77) employs exactly the explicitly relational $r_{ij}$-configurations which we have seen already (although they, somewhat arbitrarily, restrict the configurations to the embeddable ones, i.e. those where the $r_{ij}$ can be interpreted as spatial distances). 

The predictions of Newtonian gravity can thus be accounted for by an explicitly relational theory. What about the rest of our `fundamental' theories? For three hundred years after Newton's work, physics advanced largely on individualist/substantivalist lines. Several developments were pertinent to the individualist-relational question. Firstly, special relativity abolished absolute space, but replaced it with an absolute background spacetime. The terms of the substantivalist-relationalist debate have superficially changed, in that the argument is now about whether spacetime, rather than space, has an independent existence. Nevertheless, physics after special relativity was still primarily phrased in terms of individualist configurations, given by the worldlines of particles in spacetime and the values of tensor fields at each point in spacetime. Secondly, fields became a ubiquitous ingredient in physical theories. This is important for us, because it is very difficult to see how one could formulate a field theory in explicitly relational terms. Thus it is that general relativity (GR), the only one of our current `fundamental' physical theories which implements a relationalist view of spacetime, is phrased in terms of ID-relational configurations. 

This is potentially a problem for our programme. If (i) there is no obvious way to formulate a field theory in terms of explicitly relational configurations, the best we can do being ID-relational configurations; and (ii) ID-relational configurations cannot be non-embeddable; then we conclude that any attempt to develop physical theories employing non-embeddable relational configurations will require us to eschew fields. How then can we hope to recover the existing theories of classical relativistic physics, given that fields are almost ubiquitous in their formulation? 

In fact, some of classical relativistic physics has already been re-formulated without reference to fields. For example, Wheeler and Feynman proposed a field-free electrodynamics \cite{WF49,WF45} (henceforth WF-EM) which essentially reproduces the predictions of Maxwell's EM. This theory makes no fundamental reference to fields: it is stated purely in terms of the worldlines of charged particles. 
Fields can be introduced in the equations of motion as a calculational convenience, but have no independent degrees of freedom. 
The theory does not make the same predictions as Maxwell's for the motion of charged particles in general: 
\begin{enumerate}
\item
It predicts the same motion for a charged particle as would occur in the Maxwell theory if all the other charged particles were generating a 4-potential equal to half the retarded plus half the advanced Lienard-Wiechert potentials, rather than just the standard retarded potential. The advanced potentials are cause for concern because they suggest the effects of radiation will be felt before it is emitted, which is clearly in contradiction with what we see! But, as will be explained shortly, the theory can account for why we don't see such advanced action. 
\item
Furthermore, in Maxwell's theory the field can carry energy and momentum away from a particle (with consequent influence on the particles motion) regardless of whether this energy and momentum is ever absorbed by another charged particle. In contrast, in WF-EM energy and momentum cannot be lost from a charge unless they are at some point recovered by another charge. 
\end{enumerate}

However, beginning with WF-EM, one can show that in a universe consisting of a large enough number of charges, with appropriate initial conditions, standard Maxwell EM with retarded potentials is a good description of small subsystems of charges. In rough terms, advanced and retarded potentials cancel out to leave only the normal retarded potentials, plus an extra potential which exactly accounts for the loss of energy and momentum from a radiating charge. This second term which explains the radiation reaction force is the only noticeable effect of the advanced potentials. 
There is a striking similarity here with the Machian procedure for accounting for inertial motion within a relational description: in both cases the universe as a whole is described by new laws, but the predictions of the existing theory are recovered in the description of small subsystems of the universe. Note that the configurations of WF-EM are still individualist - they consist of timelike worldlines in a background Minkowski spacetime. Nonetheless this version of electrodynamics is clearly more amenable to a re-working in explicitly relational terms than is the standard Maxwell field theory. 

GR currently has no such field-free reformulation (although moves in this direction were made by Hoyle and Narlikar \cite{HNGravity}). Eliminating the metric field is slightly more awkward because of its importance in determining the causal structure of the spacetime. Initially we will concentrate solely on recovering electromagnetism, in the form of WF-EM, but will keep in mind that we ultimately want to incorporate GR as well. 
All this suggests that we should begin by looking for relational configurations some subset of which derive from individualist configurations of timelike worldlines in Minkowski space.

\section{`Relativistic' relational configurations}

We are looking for a RCS $\mathcal{R}$ such that: 
\begin{enumerate}
\item
There exists a proper subspace $\mathcal{R}_0$ of $\mathcal{R}$ whose configurations can be derived from individualist configurations $I$ of timelike worldlines in Minkowski space. The configurations $R\in\mathcal{R}_0$ should contain enough information to reconstruct all physically significant elements of $I$. 
\item
The form of the relational configurations should facilitate the statement of appropriate dynamics on $\mathcal{R}$. 
\end{enumerate}

We see that the $r_{ij}$-configurations fulfil analogous criteria if we were seeking to re-derive $\mathbf{x}_i$-configurations of particles in 3-dimensional space, moving according to Newtonian gravity . $r_{ij}$-configurations satisfying embedding constraints can be derived from $\mathbf{x}_i$-configurations and such a $r_{ij}$-configuration determines a $\mathbf{x}_i$-configuration up to translations and rigid rotations of the whole system of particles in the space, transformations which are not dynamically significant. Finally, the $r_{ij}$ seem like a good choice of relational data since they appear directly in the gravitational interaction terms of the theory we are trying to recover. 

But $r_{ij}$-configurations are inadequate for our purposes because they fail to incorporate two key insights from relativity: 
\begin{enumerate}
\item
An $r_{ij}$-configuration describes the configuration of multiple separated particles at a particular instant - thus it embodies a preferred notion of simultaneity, which is incompatible with relativistic ideas. 
\item
The fundamental degrees of freedom in an $r_{ij}$-configuration, which for embeddable configurations are the distances between particles. But according to special relativity distance is a frame-dependent notion. 
\end{enumerate}

\begin{figure}\label{fig_rij_to_2IR}
\def\svgwidth{0.9\columnwidth}
\scriptsize
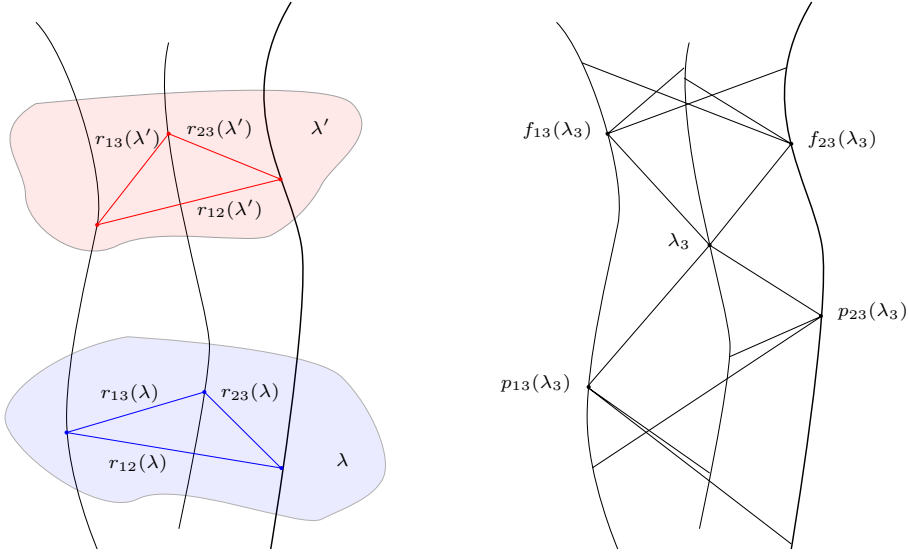
\normalsize
\caption{$r_{ij}$-configurations partition the history into slices; 2IRs do not. Only a small sample of related pairs of instants are shown in each case.}
\end{figure}

We now generalise the $r_{ij}$-configurations to eliminate these non-relativistic features. First we address the issue of absolute simultaneity. This is clarified by taking a `spacetime' view of evolving $r_{ij}$-configurations: we have $n$ particle worldlines, all of them labelled with a common parameter $\lambda$. Given two worldlines $W_i$ and $W_j$, for every pair of instants with the same value of $\lambda$ there is a distance $r_{ij}$ defined. Note that dynamical quantities are not assigned to other pairs of instants: there is no distance between the instant $\lambda$ on $W_1$ and the instant $\lambda'$ on $W_2$. We can turn this observation on its head: forget about the parameterisation of the worldlines (which is merely a convention) and note that, given the set of all instants on all worldlines, there is an equivalence relation: two points are related if they are both associated with a distance $r_{ij}$. This equivalence relation partitions the worldlines into `slices' each containing a single instant from each worldline (see figure \ref{fig_rij_to_2IR}). It is then natural to parameterise each of the worldlines so that those instants which are related have the same value of parameter, which can then be interpreted as a global time. If we want to eliminate absolute simultaneity, then the association of pairs of points with dynamical variables should not allow for such a partition into slices. 

One natural generalisation might be to relate \emph{all} pairs of points: we could then perhaps assign numbers to each pair and interpret these as spacetime intervals rather than distances. But we will take a different approach here. BB77 is designed to recover Newtonian gravity. Analogously our aim is to recover a classical relativistic theory: classical electrodynamics. The pairs of points which are related by the $r_{ij}$-configurations of BB77 have a special dynamical significance to each other in Newtonian gravity: the equations of motion relate quantities such as accelerations and positions for different particles at these instants\footnote{We can rewrite the equations of motion to relate quantities at different pairs of instants, but the equations then take on a messy particle-number-dependent and time-dependent form, so these instants do have genuine dynamical significance to one another. It is a virtue of the relational approach e.g. BB77 that relationships between other pairs of instants are not even defined.}. Are there similar, `dynamically significant' pairs of instants in classical electrodynamics? Yes, given a collection of charged particles, the acceleration of particle 1 at time $t_1$ is related, via the Lienard-Wiechert potentials\footnote{These are the electromagnetic 4-potentials due to a single point charge in arbitrary motion. See for example \cite{Jackson} for details.}, to the positions and velocities of all other particles \emph{at the instants at which their worldlines intersect the past light-cone of particle 1 at $t_1$}. This is made even clearer in WF-EM where all transfers of energy and momentum occur between worldline points connected by null geodesics. This pairing of instants by dynamics in electromagnetism is not an equivalence relation - it is clearly not transitive. Thus the dynamically connected instants are not partitioned into disjoint `slices' (again see figure \ref{fig_rij_to_2IR}). This is the fundamental reason why there is no absolute simultaneity in special relativity. Instead a complicated web of connected instants is formed. Inspired by this analogy we propose that every instant $\lambda_i$ on a worldline $W_i$ is related to two instants, denoted $f_{ji}(\lambda_i)$ and $p_{ji}(\lambda_i)$, on each other worldline $W_j$. We will call this collection of functions a \emph{two instant relation} or \emph{2IR}. 

The `spacetime' version of the $r_{ij}(\lambda)$-configurations related pairs of instants on different worldlines, and then associated a quantity $r_{ij}$ with each pair. So far we have related pairs of instants, so to complete the analogy we should associate some quantity with these pairs. In the $r_{ij}$ case any number associated with the paired instants has a natural interpretation in embeddable configurations: as an interparticle distance. In the 2IR case, a number associated with paired instants has no obvious interpretation for embeddable configurations. When embedded these instants are supposed to be connected by null geodesics, so the spacetime interval for all these pairs is zero. Arguably the most dynamically significant quantity is the 3-space distance between the two instants, since this appears in the Lienard-Wiechert potentials, but this is a frame-dependent quantity. 

\begin{figure}\label{fig_rel_distance}
\def\svgwidth{0.4\columnwidth}
\scriptsize
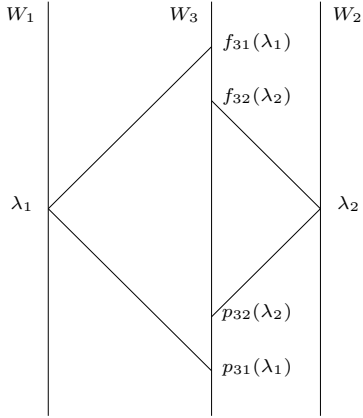
\normalsize
\caption{The 2IR allows for a notion of relative distance from a reference worldline: $W_2$ is `closer' to $W_3$ than $W_1$ is.}
\end{figure}

Perhaps we could dispense with numbers, and take the `undecorated' 2IRs as our relational configurations. In this case the only real structure would be the pattern of connections between different worldlines. 
In the $r_{ij}$ case dropping the numbers and retaining only the structure of connected instants would leave a totally uninteresting structure which was essentially the same at all times. But the 2IR has potential for much more complex structure. 
There is certainly some conceptual appeal to the idea of recovering spacetime from an entirely `numberless' structure. The numbers we attribute to space and time (distances, durations, etc.) typically actually represent structural relationships between material bodies; for example, the distance between two bodies is really a measure of how many times you can fit a reference object (a measuring rod) in between them. The 2IR does seem to have potential in this regard. If $s_{ji}(\lambda_i)\subset W_j$ is the subset of worldline $W_j$ between the points $f_{ji}(\lambda_i)$ and $p_{ji}(\lambda_i)$, then the overlaps of $s_{31}$ and $s_{32}$ give us some notion of the relative distance of $W_1$ and $W_2$ from $W_3$, as shown in figure \ref{fig_rel_distance}. 

However, one of our requirements for our RCS was that we should be able to reconstruct all physically significant elements of the spacetime configurations from the relational configurations. In fact, a globally scaled configuration of worldlines in Minkowski space would give rise to the same 2IR as the original configuration, so the global scale cannot be reconstructed from the 2IR. Classical electrodynamics is not scale-invariant: given a configuration of worldlines which is a solution to the electromagnetic equations of motion, applying a global scaling factor would result in a configuration of worldlines which was not a solution. 
In these individualist configurations, it's the metric of the spacetime that provides the scale, but the observable consequences of a lack of scale invariance are that different characteristic structures of material bodies have different relative sizes. This suggests that, beginning from relational configurations, a privileged scale factor might emerge for small subsystems in a `Machian' fashion, due to structural features of the configuration of the rest of the universe. If this is the case (and we would need to show exactly how it worked), we need not worry about introducing scale directly into our relational configurations. 
If we did want to introduce scale directly into the relational configurations, an alternative to attaching numbers to paired instants is to apply a metric to each worldline, which would be interpretable as proper time for embedded configurations. This would remove the scale ambiguity in embedding. We will refer to such relational configurations as \emph{two-instant relations with proper times} or \emph{$\tau$-2IR}s. Most of our subsequent analysis will apply to the simpler 2IRs without proper times, but we will mention where the analysis might differ for $\tau$-2IRs.

\section{Two instant relations - 2IRs}

We first define the two-instant relation precisely. 

\begin{definition}
A \emph{worldline} is a manifold homeomorphic with $\mathbb{R}$, with the total ordering compatible with $\mathbb{R}$ defined on it. 
\end{definition}

\begin{definition}
Consider a collection of $n$ worldlines $\{W_i\}$. For each pair of worldlines $W_i$ and $W_j$, we define a pair of \emph{continuous}, \emph{strictly monotonic} functions $f_{ij},p_{ij}:W_j\rightarrow W_i$ called the \emph{future} and \emph{past} functions respectively, which satisfy: 
\begin{equation} f_{ij}(\lambda)>p_{ij}(\lambda)\end{equation}
\begin{equation} f_{ij} = p^{-1}_{ji}\end{equation}
We term this collection of functions the \emph{two-instant relation (2IR)} on the worldlines. 
\end{definition}

In rough terms, every point on some worldline is related to two points on each other worldline. Note that any continuous strictly monotonic function (like $f_{ij}$ and $p_{ij}$) is also bijective. Almost invariably we will apply coordinates to each worldline: 
\begin{definition}
A \emph{worldline parameter} for a worldline $W$ is a continuous monotonic function $\lambda: W\rightarrow \mathbb{R}$. 
\end{definition}

The functions $f_{ij}$ and $p_{ij}$ are maps between worldlines. However, to do any useful calculations we really need to deal with coordinate-dependent representations of these functions. 
\begin{definition}
Given an $n$-worldline 2IR, and worldline parameters $\lambda_i: W_i\rightarrow \mathbb{R}$, the \emph{parameterised representations} of the 2IR functions are the following functions:
\begin{equation} 
F_{ij} = (\lambda_i\circ f_{ij}\circ\lambda_j^{-1}) \;,\; P_{ij} = (\lambda_i\circ p_{ij}\circ\lambda_j^{-1}):\mathbb{R}\rightarrow\mathbb{R}
\end{equation}
\end{definition}

Note that any two parameterisations $\lambda'_i$ and $\lambda_i$ of a worldline $W_i$ will be related by a continuous monotonic function $r_i := \lambda_i'\circ\lambda_i^{-1}$: i.e. $\lambda'_i=r_i\circ\lambda_i:W_i\rightarrow \mathbb{R}$. Applying coordinate re-parameterisations to the worldlines of a 2IR will result in parameterised representatives with different functional forms which nonetheless correspond to the same 2IR: $F'_{ij}=r_i\circ F_{ij}\circ r_j^{-1}$ and $P'_{ij}=r_i\circ P_{ij}\circ r_j^{-1}$. Henceforth we will deal almost exclusively with the coordinate representatives $F_{ij}$ and $P_{ij}$, recalling that many different such representatives may correspond to the same 2IR.\footnote{In the case of a $\tau$-2IR the distance functions on the worldlines clearly suggest a preferred parameterisation.} 

\subsection{Equivalence and classification of 2IRs}\label{sec_2IR_equivalence}

In the 2IR formalism, a point on a worldline is distinguished by the points on other worldlines to which it is related; it has no other intrinsic properties. Thus it seems reasonable that any two 2IRs which have the same pattern of relationships amongst the instants on their worldlines should be viewed as equivalent. The following definition makes this precise: 
\begin{definition}
Two 2IRs $\{W_i, f_{ij}, p_{ij}\}$ and $\{W_i, f'_{ij}, p'_{ij}\}$ are \emph{equivalent} if there exist continuous monotonic maps $s_i: W_i\rightarrow W_i$ such that $f'_{ij} = s_i\circ f_{ij}\circ s^{-1}_j$ and similarly for $p_{ij}$ and $p'_{ij}$
\footnote{Equivalence for $\tau$-2IRs is stricter: the functions $s_i$ must also preserve distances on the worldlines. Most of the results in the remainder of this section do not apply to $\tau$-2IRs.}.
\end{definition}
Here we have viewed the transformation in active terms. With the introduction of coordinates we can get a passive view of equivalence. 
\begin{lemma}
Two 2IRs are equivalent iff their sets of parameterised representations $\{F_{ij},P_{ij}: \mathbb{R}\rightarrow\mathbb{R}\}$ are identical. 
\end{lemma}

\textbf{Proof:} 
Consider two 2IRs, $R = \{W_i, f_{ij}, p_{ij}\}$ and $R' = \{W'_i, f'_{ij}, p'_{ij}\}$. Suppose we can find parameterisations for them, $\lambda_i$ and $\lambda'_i$ respectively, such that the parameterised representations are equal: $\lambda_i\circ f_{ij}\circ\lambda_j^{-1} = F_{ij} = \lambda'_i\circ f'_{ij}\circ\lambda'^{-1}_j$. Then we have $f'_{ij} = (\lambda'^{-1}_i\circ\lambda_i)\circ f_{ij}\circ(\lambda'^{-1}_j\circ\lambda_j)^{-1}$. If we define $s_i = (\lambda'^{-1}_i\circ\lambda_i): W_i\rightarrow W'_i$ which is a continuous monotonic function, then we have $f'_{ij} = s_i\circ f_{ij}\circ s_j^{-1}$. A similar argument applies to $p_{ij}$ and $p'_{ij}$, and thus $\mathcal{R}$ and $\mathcal{R}'$ are equivalent.
Conversely, assume $R$ and $R'$ are equivalent, i.e. $f_{ij} = s_i \circ f_{ij}' \circ s^{-1}_j$. Choose a set of worldline parameters $\{\lambda_i: W_i\rightarrow\mathbb{R}\}$, giving us parameterised representations $F_{ij} = \lambda_i \circ f_{ij} \circ \lambda^{-1}_j$ for $f_{ij}$. We can write $F_{ij} = \lambda_i \circ (s_i \circ f_{ij}' \circ s^{-1}_j) \circ \lambda^{-1}_j = (\lambda_i \circ s_i) \circ f_{ij}' \circ (s^{-1}_j \circ \lambda^{-1}_j)$. Now $\lambda_i \circ s_i: W_i\rightarrow \mathbb{R}$ are a valid set of worldline parameters and thus $F_{ij}$ is also a parameterised representation of $f_{ij}'$. The argument runs equally well in reverse, and for $p_{ij}$ and $p'_{ij}$. 
$\square$

So, from the passive viewpoint, one 2IR is equivalent to another if we can re-parameterise their worldlines to bring the parameterised representations of their future and past functions into the same functional form. 

Clearly, any 2IR is equivalent to many others: begin with a parameterised representation of the 2IR $\mathcal{R}$, perform a re-parameterisation of its worldlines to get a new parameterised representation, and then view this as the representation of a different 2IR $\mathcal{R}'$, in the original coordinates. The question then arises as to whether there are any inequivalent 2IRs, and if so, how to classify them. 

\begin{lemma}\label{lem_2wl_equiv}
All 2-worldline 2IRs are equivalent. 
\end{lemma}

\textbf{Sketch of proof}

First we want to find a reparameterisation $r$ of the $\lambda_2$ axis such that the transformed 2IR functions are a constant vertical distance apart. We do this as follows. 
We partition the $\lambda_1$ axis into intervals $\Lambda_1^i = ((F_{12}\circ F_{21})^{i-1}(0), (F_{12}\circ F_{21})^{i}(0)]$ and the $\lambda_2$ axis into regions $\Lambda_2^i = (F_{21}\circ(F_{12}\circ F_{21})^{i-1}(0), F_{21}\circ(F_{12}\circ F_{21})^{i}(0)]$. These intervals are shown on the diagram below. We construct $r_i$ on each $\Lambda_2^i$ successively. $r_1$ is defined on $\Lambda_2^1$ such that this interval is `stretched' or `compressed' so that $r_1\circ F_{21}$ on $\Lambda_1^1$ is a constant distance $2a$ above $P_{21}$. 

$P_{21}$ on $\Lambda_1^1$ is unaffected by this change, but it is affected on $\Lambda^2_1$. So when defining $r_2$ on $\Lambda_2^2$ we require that $r_2\circ F_{21}=r_1\circ P_{21} + 2a$ on $\Lambda_1^2$. We continue this inductive process, so that $r_i$ is defined on $\Lambda_2^i$ such that $r_i\circ F_{21} = r_{i-1}\circ P_{21} + 2a$ on $\Lambda_1^i$. A similar process with the roles of $F_{21}$ and $P_{21}$ interchanged is used to define $r_i$ on $\Lambda_2^i$ for negative $i$. 
This may not be an efficient way to calculate $r$ but it is clear that this method yields $r$ such that $r\circ F_{21} - r\circ P_{21}$ is a constant along the entire $\lambda_1$ axis. 

We then turn to the reparameterisation $s$ of the $\lambda_1$ axis. Note firstly $F_{21}\circ s - P_{21}\circ s = F_{21} - P_{21}$ i.e. rescaling the $\lambda_1$ axis has no effect on the vertical distance between the 2IR functions, and specifically if they are a constant distance apart before the application of $s$, they will be after as well. Note secondly that if we have a free choice of $s$ we can bring one of the functions into any form we please. Specifically given any $P_{21}$ we can find an $s$ s.t. $P_{21}\circ s = P'_{21}$: we just choose $s = P^{-1}_{21}\circ P'_{21}$ which is continuous and monotonic, as required. 

Thus, given any $F_{21}, P_{21}$, we choose $r$ s.t. the resulting functions $F''_{21} = r\circ F_{21}$ and $P''_{21} = r\circ P_{21}$ satisfy $F''_{21} = P''_{21} + 2a$, and then choose $s = P''^{-1}_{21} \circ P'_{21}$ where $P'_{21}(\lambda_1) = \lambda_1 -a$. We then have $F'_{21}(\lambda_1) = F''_{21}\circ s (\lambda_1) = \lambda_1 +a$ and $P'_{21}(\lambda_1) = P''_{21}\circ s (\lambda_1) = \lambda_1 -a$. Having established that any 2IR can be brought into this form, we have established that they are all equivalent\footnote{Readers may be concerned that for smooth functions $F_{21}$ and $P_{21}$, $r\circ F_{21}$ and $r\circ P_{21}$ will not be smooth in general. Our definition of the 2IR did not require $f_{ij}$ and $p_{ij}$ to be smooth, only continuous and monotonic. It is not currently known whether this equivalence result on 2-worldline 2IRs would survive, if smoothness was enforced for 2IR functions and reparameterisations.}. $\square$

\begin{figure}
\def\svgwidth{0.9\columnwidth}
\small
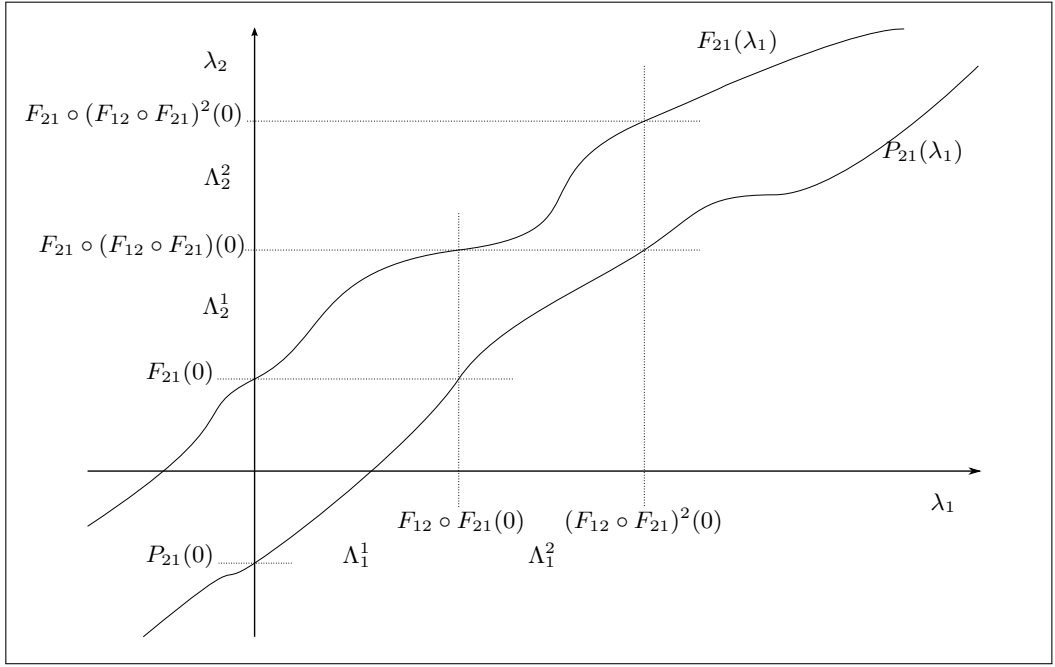
\normalsize
\caption{Diagram for proof of 2-worldline 2IR equivalence}
\end{figure}

\begin{lemma}
An $n$-worldline 2IR is characterised at most by $n(n-2)$ continuous monotonic functions. 
\end{lemma}
\textbf{Proof}: Label the worldlines with $i = 1\dots n$. Consider first just worldlines $W_1$ and $W_2$. Associated with these worldlines are two continuous monotonic functions $F_{21}$ and $P_{21}$, but as shown in lemma \ref{lem_2wl_equiv}, re-parameterisations of $W_1$ and $W_2$ allow these functions to be put into any form. Now also consider worldline $W_3$. There are now an additional four continuous monotonic functions to consider ($F_{31}, P_{31}, F_{32}, P_{32})$. Reparameterising $W_3$ allows us to bring one of these functions into any form we like, leaving three continuous monotonic functions to characterise the 2IR. We continue to add in worldlines: on the addition of $W_k$ we must consider an additional $2(k-1)$ 2IR functions, one of which can be brought into any form by re-parameterising $W_i$. There are $n(n-1)$ 2IR functions in total, and we have seen that $n$ of them can be transformed arbitrarily via reparameterisations of the $n$ worldlines. Thus each 2IR is characterised at most by $n(n-1) -n = n(n-2)$ continuous monotonic functions. 
$\square$

\begin{figure}\label{fig_3wl_inequiv}
\def\svgwidth{0.8\columnwidth}
\begin{center}
\scriptsize
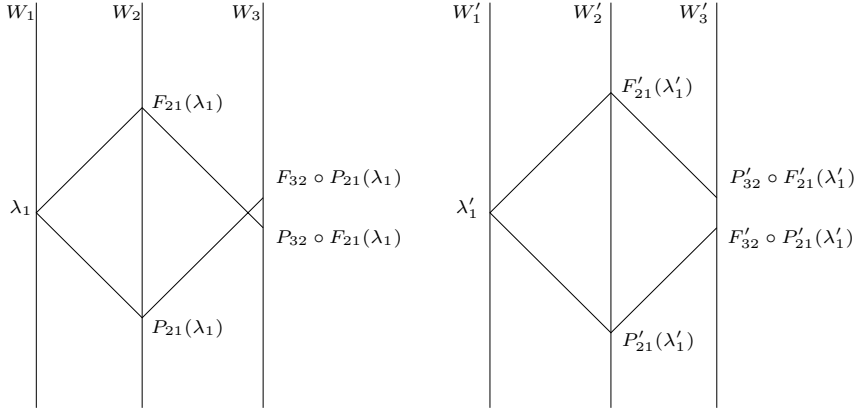
\normalsize
\end{center}
\caption{No equivalence could map $\lambda_1$ to $\lambda'_1$}
\end{figure}

It remains an open question whether the degrees of freedom characterising a 2IR can be reduced any further - it seems unlikely, but we have no explicit proof. Certainly it is clear that for three worldlines or more, there are inequivalent 2IRs. For example consider a 3-worldline 2IR $\mathcal{R}$ such that for all instants $\lambda_1$ on $W_1$, $F_{32}\circ P_{21}(\lambda_1) < P_{32}\circ F_{21}(\lambda_1)$. This clearly cannot be equivalent to a 3-worldline 2IR $\mathcal{R}'$ such that for all instants $\lambda'_1$ on $W'_1$, $F'_{32}\circ P'_{21}(\lambda'_1) > P'_{32}\circ F'_{21}(\lambda_1)$ (see figure \ref{fig_3wl_inequiv}).

\subsection{Embeddibility of 2IRs}\label{sec_2IR_embed}

Given a configuration of timelike worldlines $\{W_i\}$ in Minkowski spacetime the \emph{derived 2IR} is defined as follows\footnote{For a $\tau$-2IR we must supplement this definition by saying that the distance functions on the worldlines are given by taking the difference of the proper times of any pair of instants. Once again, many results in this section will not apply to $\tau$-2IRs: typically their embeddibility is subject to more stringent conditions.}: 

\emph{If $\lambda\in W_i$ is a point on the $i^\text{th}$ worldline, $f_{ji}(\lambda)\in W_j$ is the intersection of the future light-cone of $\lambda$ with $W_j$;  $p_{ji}(\lambda)\in W_j$ is the intersection of the past light-cone of $\lambda$ with $W_j$.}

In accordance with section \ref{sec_rel_ind} a 2IR is \emph{embeddable} if it can be derived in this fashion from a configuration of timelike worldlines in Minkowski space. Two questions now arise: 
\begin{enumerate}
\item \textbf{Possibility of embedding}: Are all 2IRs embeddable? If not, how do we decide whether a given 2IR is embeddable? 
\item \textbf{Uniqueness of embedding}: If a 2IR is embeddable, is the embedding unique? Put another way, how much does the 2IR tell us about the embedding? 
\end{enumerate}
For the 2-worldline case there is only one 2IR and these questions are easily answered: (1) the 2-worldline 2IR is embeddable; (2) it embeds totally non-uniquely as \emph{any} configuration of two timelike worldlines i.e. the 2IR tells us nothing about the embedding. For $n>2$, the situation is much more complicated. 

If a 2IR is embeddable, it implies the existence of four \emph{embedding functions} defined for each worldline $x_i^\mu: W_i \rightarrow \mathbb{R}$, which furthermore satisfy a family of conditions (which we will term the \emph{light-cone conditions} or \emph{LC conditions}), formalising the idea that worldline instants related by 2IR functions are connected by null geodesics when embedded in the spacetime. 
In the case of inertial coordinates we have, for all $W_i,W_j, \lambda_j\in W_j$: 
\begin{equation}\label{eq_F_embed}
[t_i\circ f_{ij}(\lambda_j) - t_j(\lambda_j)]^2 = [x_i\circ f_{ij}(\lambda_j) - x_j(\lambda_j)]^2 + [y_i\circ f_{ij}(\lambda_j) - y_j(\lambda_j)]^2 + [z_i\circ f_{ij}(\lambda_j) - z_j(\lambda_j)]^2
\end{equation}
\begin{equation}\label{eq_P_embed}
[t_j(\lambda_j)-t_i\circ p_{ij}(\lambda_j)]^2 = [x_i\circ p_{ij}(\lambda_j) - x_j(\lambda_j)]^2 + [y_i\circ p_{ij}(\lambda_j) - y_j(\lambda_j)]^2 + [z_i\circ p_{ij}(\lambda_j) - z_j(\lambda_j)]^2
\end{equation}
Each $\lambda_j\in W_j$ appears in $2n-2$ light-cone conditions, one for each instant on other worldlines to which it is related. 
The question of whether a 2IR is embeddable comes down to whether the corresponding LC conditions are consistent i.e. are there any choices of embedding functions which satisfy them all. The question of how unique the embedding is would be answered by solving these constraints to get some family of sets of embedding functions consistent with the 2IR. These problems are not straightforward because of the complexity of the couplings between the LC conditions. We are trying to constrain $4n$ \emph{functional} degrees of freedom (the embedding functions). There are $2n-2$ LC conditions for \emph{every instant} on every worldline, thus we have a continuous infinity of constraints. Furthermore, these constraints are coupled together in a non-trivial way. Each constraint corresponds to a pair of points which are connected by the 2IR. Each of these two points is then connected to a further $2n-3$ points, and each of these connections corresponds to a constraint which is coupled with our original constraint. Thus every constraint is coupled to another $4n-6$ constraints, to form a complicated web of couplings. 

Ideally we would like some algorithm which, when supplied with a 2IR, first determined if it was embeddable, and if so, determined which embeddings were possible. We are still some way off having this, but we do have some preliminary results relating to embeddibility. Firstly, a condition which must be satisfied for embeddibility to be possible: 

\begin{lemma}\label{lem_local_cond}
If a 2IR $\mathcal{R}$ is embeddable, the following conditions hold for every three distinct worldlines $(W_i, W_j,W_k)$ in $\mathcal{R}$:
\begin{equation}\label{eq_PP<P}
p_{kj} \circ p_{ji} \leq p_{ki}
\end{equation}
\begin{equation}\label{eq_FP>P}
f_{kj} \circ p_{ji} \geq p_{ki}
\end{equation}
\begin{equation}\label{eq_PF>P}
p_{kj} \circ f_{ji} \geq p_{ki} 
\end{equation}
Similar equations hold for (i) cyclic permutations of $i, j$ and $k$, and (ii) swapping $p$ and $f$ functions and reversing the inequality. 
These conditions are termed the \emph{locality conditions}. 
\end{lemma}

\begin{figure}\label{fig_pp<p_proof}
\def\svgwidth{0.8\columnwidth}
\begin{center}
\scriptsize
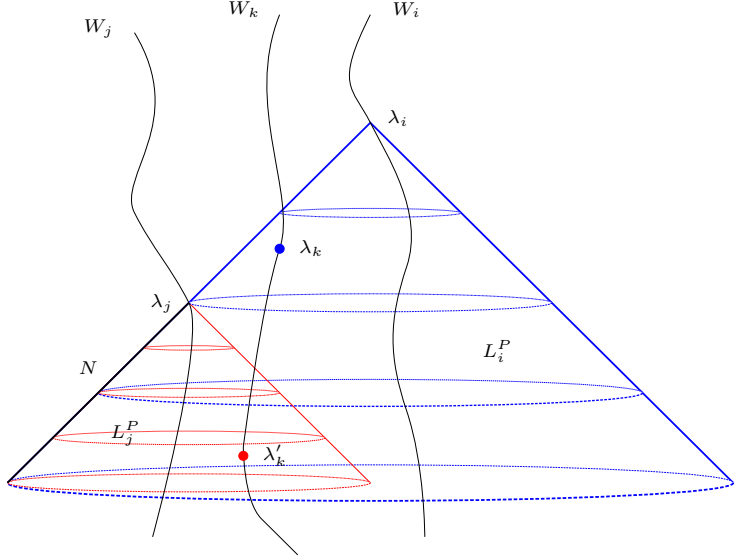
\normalsize
\end{center}
\caption{$p_{kj} \circ p_{ji} \leq p_{ki}$ for embeddable 2IRs}
\end{figure}

\textbf{Proof:} Consider $\lambda_i \in W_i$, and the point in spacetime to which it corresponds $\sigma_i(\lambda_i)$. Denote the past light-cone of $\sigma_i(\lambda_i)$ by $L^P_i$ (i.e. the collection of spacetime points connected to $\sigma_i(\lambda_i)$ by null geodesics, and in the past of $\sigma_i(\lambda_i)$). Suppose $\lambda_j = p_{ji}(\lambda_i)$. Then $\sigma_j(\lambda_j) \in L^P_i$. Now denote the past light-cone of $\sigma_j(\lambda_j)$ by $L^P_j$. $L^P_i$ and $L^P_j$ share points along a null geodesic $N$ which terminates at $\sigma_j(\lambda_j)$. Otherwise the points of $L^P_j$ are entirely `enclosed' within those of $L^P_i$, in the sense that any timelike geodesic that intersected $L^P_j$ would have to \emph{later} intersect $L^P_i$ (see figure \ref{fig_pp<p_proof}). 
Now consider two points $\lambda_k = p_{ki}(\lambda_i)$ and $\lambda'_k = p_{kj}(\lambda_j)$. We have $\lambda_k, \lambda'_k\in W_k$ which is mapped by $\sigma_k$ into a timelike geodesic in the spacetime, and $\sigma_k(\lambda_k) \in L^P_i$ and $\sigma_k(\lambda'_k) \in L^P_j$. 
Thus by the argument above we have $\lambda'_k < \lambda_k$, or equivalently $p_{kj}\circ p_{ji} (\lambda_i) \leq p_{ki}(\lambda_i)$ with equality holding only if $\lambda_k \in N$. This holds $\forall \lambda_i\in W_i$ so we have proved (\ref{eq_PP<P}).  
To prove (\ref{eq_FP>P}) we apply $f_{jk}$ to both sides of (\ref{eq_PP<P}), and a similar manoeuvre yields (\ref{eq_PF>P}). 
$\square$

Any 2IR which fails to satisfy these conditions at all points of all its worldlines will not be embeddable. In fact some of these 2IRs display a more extreme variant of non-embeddibility, somewhat akin to Escher's endless staircase: 
\begin{center}
\includegraphics[scale=0.5]{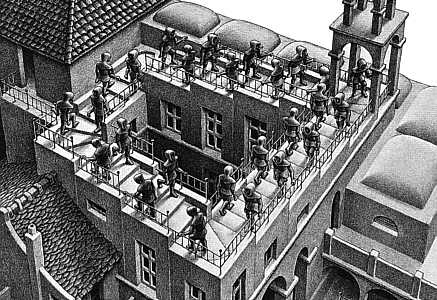}
\end{center}
\begin{definition}\label{def_Escher}
The \emph{Escher conditions} are: 
\begin{equation}\label{eq_FF>P}
f_{kj} \circ f_{ji} > p_{ki}
\end{equation}
\begin{equation}\label{eq_PP<F}
p_{kj} \circ p_{ji} < f_{ki}
\end{equation}
\end{definition}
Suppose for example that the first condition is not satisfied i.e. $f_{kj} \circ f_{ji} < p_{ki}$. Apply $f_{ik}$ to both sides to yield $f_{ik} \circ f_{kj} \circ f_{ji} (\lambda) < \lambda$. In this case following three future pointing geodesics from $\lambda\in W_i$ brings us back to $W_i$ at a point in the past of our original point! 

The next result shows that some details of the embedding can be directly inferred from the 2IR. Consider the locality conditions from lemma \ref{lem_local_cond}. These are inequalities -- what happens when we have equality in one of these conditions? For example consider $p_{kj} \circ p_{ji} = p_{ki}$. Referring to figure \ref{fig_pp<p_proof}, we see that this corresponds to $\lambda_k = \lambda'_k$, and it is clear that in this case $\lambda_k$ must like on the null geodesic $N$, since this is the only point where the two light-cones coincide. Thus, equality in this locality condition implies collinearity when the 2IR is embedded, and the same is easily seen to be true for all the other locality conditions. 

\begin{lemma}\label{lem_collinearity}
Consider a 2IR $\mathcal{R}$ with three or more worldlines, and a point $\lambda_i$ on a worldline $W_i$. If when applying both sides of one of the locality conditions to $\lambda_i$ the condition is satisfied \emph{with equality}, then in any embedding of $\mathcal{R}$, the three spacetime points corresponding to $\lambda_i$, the LHS of the condition applied to $\lambda_i$, and the RHS of the condition applied to $\lambda_i$, will be collinear in spacetime. Specifically we have: 
\def\svgwidth{0.8\columnwidth}
\begin{center}
\scriptsize
\begingroup%
  \makeatletter%
  \providecommand\color[2][]{%
    \errmessage{(Inkscape) Color is used for the text in Inkscape, but the package 'color.sty' is not loaded}%
    \renewcommand\color[2][]{}%
  }%
  \providecommand\transparent[1]{%
    \errmessage{(Inkscape) Transparency is used (non-zero) for the text in Inkscape, but the package 'transparent.sty' is not loaded}%
    \renewcommand\transparent[1]{}%
  }%
  \providecommand\rotatebox[2]{#2}%
  \ifx\svgwidth\undefined%
    \setlength{\unitlength}{800bp}%
    \ifx\svgscale\undefined%
      \relax%
    \else%
      \setlength{\unitlength}{\unitlength * \real{\svgscale}}%
    \fi%
  \else%
    \setlength{\unitlength}{\svgwidth}%
  \fi%
  \global\let\svgwidth\undefined%
  \global\let\svgscale\undefined%
  \makeatother%
  \begin{picture}(1,0.38)%
    \put(0,0){\includegraphics[width=\unitlength]{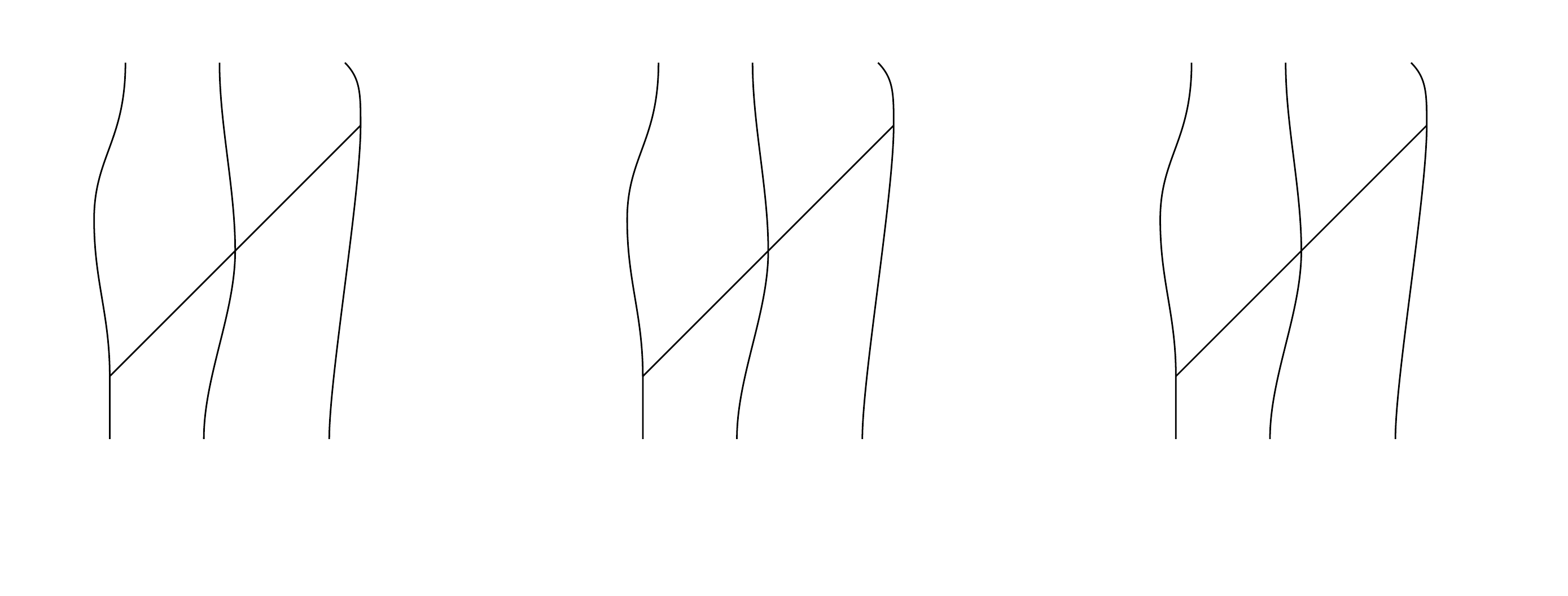}}%
    \put(0.08,0.36){\color[rgb]{0,0,0}\makebox(0,0)[b]{\smash{\small{$W_i$}}}}%
    \put(0.14,0.36){\color[rgb]{0,0,0}\makebox(0,0)[b]{\smash{\small{$W_k$}}}}%
    \put(0.22,0.36){\color[rgb]{0,0,0}\makebox(0,0)[b]{\smash{\small{$W_j$}}}}%
    \put(0.05,0.14){\color[rgb]{0,0,0}\makebox(0,0)[b]{\smash{$\lambda_i$}}}%
    \put(0.18,0.22){\color[rgb]{0,0,0}\makebox(0,0)[b]{\smash{$\lambda_k$}}}%
    \put(0.2514108,0.29886584){\color[rgb]{0,0,0}\makebox(0,0)[b]{\smash{$\lambda_j$}}}%
    \put(0.13,0.06){\color[rgb]{0,0,0}\makebox(0,0)[b]{\smash{\normalsize{$p_{ij}(\lambda_j) = p_{ik}\circ p_{kj}(\lambda_j)$}}}}%
    \put(0.13,0.02){\color[rgb]{0,0,0}\makebox(0,0)[b]{\smash{\normalsize{$f_{ji}(\lambda_i) = f_{jk}\circ f_{ki}(\lambda_i)$}}}}%
    \put(0.42,0.36){\color[rgb]{0,0,0}\makebox(0,0)[b]{\smash{\small{$W_i$}}}}%
    \put(0.48,0.36){\color[rgb]{0,0,0}\makebox(0,0)[b]{\smash{\small{$W_j$}}}}%
    \put(0.56,0.36){\color[rgb]{0,0,0}\makebox(0,0)[b]{\smash{\small{$W_k$}}}}%
    \put(0.39,0.14){\color[rgb]{0,0,0}\makebox(0,0)[b]{\smash{$\lambda_i$}}}%
    \put(0.52,0.22){\color[rgb]{0,0,0}\makebox(0,0)[b]{\smash{$\lambda_j$}}}%
    \put(0.59141077,0.29886584){\color[rgb]{0,0,0}\makebox(0,0)[b]{\smash{$\lambda_k$}}}%
    \put(0.47,0.06){\color[rgb]{0,0,0}\makebox(0,0)[b]{\smash{\normalsize{$p_{ij}(\lambda_j) = p_{ik}\circ f_{kj}(\lambda_j)$}}}}%
    \put(0.47,0.02){\color[rgb]{0,0,0}\makebox(0,0)[b]{\smash{\normalsize{$f_{ji}(\lambda_i) = p_{jk}\circ f_{ki}(\lambda_i)$}}}}%
    \put(0.76,0.36){\color[rgb]{0,0,0}\makebox(0,0)[b]{\smash{\small{$W_k$}}}}%
    \put(0.82,0.36){\color[rgb]{0,0,0}\makebox(0,0)[b]{\smash{\small{$W_i$}}}}%
    \put(0.9,0.36){\color[rgb]{0,0,0}\makebox(0,0)[b]{\smash{\small{$W_j$}}}}%
    \put(0.73,0.14){\color[rgb]{0,0,0}\makebox(0,0)[b]{\smash{$\lambda_k$}}}%
    \put(0.86,0.22){\color[rgb]{0,0,0}\makebox(0,0)[b]{\smash{$\lambda_i$}}}%
    \put(0.93141077,0.29886584){\color[rgb]{0,0,0}\makebox(0,0)[b]{\smash{$\lambda_j$}}}%
    \put(0.81,0.06){\color[rgb]{0,0,0}\makebox(0,0)[b]{\smash{\normalsize{$p_{ij}(\lambda_j) = f_{ik}\circ p_{kj}(\lambda_j)$}}}}%
    \put(0.81,0.02){\color[rgb]{0,0,0}\makebox(0,0)[b]{\smash{\normalsize{$f_{ji}(\lambda_i) = f_{jk}\circ p_{ki}(\lambda_i)$}}}}%
  \end{picture}%
\endgroup%

\normalsize
\end{center}
\end{lemma}

Finally we take some tentative steps towards an algorithm for determining embeddings. We consider the case of a 2-worldline 2IR. Recall the embedding constraints, equations \ref{eq_F_embed} and \ref{eq_P_embed} and suppose that we fix the embedding functions $t_1(\lambda_1)$, $x_1(\lambda_1)\dots$ etc. for worldline $W_1$, leaving us to solve for $t_2(\lambda_2)$, $x_2(\lambda_2)\dots$ etc. The ladder of coupled embedding conditions decouples into a family, indexed by $\lambda_2$, of pairs of equations (one involving $f_{12}$ the other involving $p_{12}$) for four unknowns ($t_2(\lambda_2)$, $x_2(\lambda_2)\dots$). Each pair of equations restricts the embedding of $\lambda_2$ to a 2-dimensional surface in the spacetime
\footnote{Earlier we claimed that all pairs of timelike worldlines in spacetime give rise to the same 2IR up to equivalence. So how can the 2IR constrain the embedding of the second worldline? Note that a worldline in a spacetime $M$ is a one dimensional subspace $M_1$ of the spacetime. 
Given a 2IR worldline $W$ there a continuos family of embeddings $\sigma:W\rightarrow\ M$ such that $\text{Image}(\sigma) = M_1$, all related by continuous monotonic maps. Different choices of $\sigma$ will constrain the embedding of the second worldline in different ways. }
: conceptually, to find where $\lambda_2$ should be embedded, we extend the past light-cone of $f_{12}(\lambda_2)$ and the future light-cone of $p_{12}(\lambda_2)$, and place $\lambda_2$ somewhere on the surface where they intersect. If we suppressed one spatial dimension, intuitively it's clear that this surface would be an ellipse in a titled plane. With all three spatial dimensions we have an ellipsoid as we now see.  

Without loss of generality we assume that $p_{12}(\lambda_2)$ is embedded at $t_1=x_1=y_1=z_1=0$ and $f_{12}(\lambda_2)$ is embedded at $t_1 = t_F$, $ x_1 = x_F$, $y_1=z_1=0$, such that $x_F/t_F <1$ (since the embedding of worldline 1 must be timelike). Our embedding conditions then become: 
\begin{equation}
(t_2(\lambda_2)-t_F)^2 = (x_2(\lambda_2)-x_F)^2 + y_2(\lambda_2)^2 + z_2(\lambda_2)^2
\end{equation}
\begin{equation}\label{ellipsoid_past}
t_2(\lambda_2)^2 = x_2(\lambda_2)^2 + y_2(\lambda_2)^2 + z_2(\lambda_2)^2
\end{equation}

First eliminate $y_2$ and $z_2$: 
\begin{equation}\label{restriction_to_plane}
\frac{1}{2}(t_F^2-x_F^2) = t_F t_2(\lambda_2) - x_F x_2(\lambda_2)
\end{equation}
This is the equation of a line in the $(x$-$t)$-plane, or a hypersurface in the full spacetime. Now substituting this into equation \ref{ellipsoid_past} to get a quadratic in $x_2$, and completing the square we get: 
\begin{equation}
\frac{(x_2(\lambda_2) - \frac{1}{2}x_F)^2}{t_F^2} + \frac{y_2(\lambda_2)^2 + z_2(\lambda_2)^2}{t_F^2+x_F^2} = \frac{1}{4}
\end{equation}
which is the equation of an ellipsoid in $(xyz)$-space centred on $(x_F/2,0,0)$. Thus $\lambda_2$ must be embedded somewhere on the intersection of these two surfaces, which is a 2-dimensional ellipsoid. 

It seems then that the problem of determining embedding for $n$-worldline 2IRs (with $n\geq 3$) will come down to determining whether certain families of 2-dimensional ellipsoids overlap in Minkowski spacetime. We are currently pursuing this approach.

\section{3-worldline 2IRs}\label{sec_3wl}

We have seen that the 2-worldline case is not very interesting -- there is only one 2IR, and it is embeddable in multiple ways, as any 2-worldline configuration in spacetime. With an extra worldline, things are more interesting. We have many inequivalent 2IRs. Many are not embeddable. Those that are embeddable have their embeddings strongly constrained by the details of the 2IR functions. We now look in more detail at 3-worldline 2IRs, particularly considering issues of equivalence and embeddibility. 

\subsection{The history space}\label{subsec_history_space}

A useful concept is the history space $\mathcal{H}$ of a 2IR $\mathcal{R}$. This is the Cartesian product of all of the worldlines. In the case of two worldlines $\mathcal{H}=W_1\times W_2$. A point in the history space is a pair $(\lambda_1,\lambda_2)$ where $\lambda_1$ is a point on worldline $W_1$ and $\lambda_2$ is a point on worldline $W_2$. The functions\footnote{Throughout this section we will use the parameterised representatives $F_{ij}$ and $P_{ij}$ of the 2IR functions, since we will be performing more explicit calculations, and visualising the history space} $F_{21}$ and $P_{21}$ partition the history space into three regions, as shown in figure :  
\begin{itemize}
\item $(\lambda_1,\lambda_2)\in F^2_1$ if $\lambda_2 > F_{21}(\lambda_1)$; intuitively $\lambda_2$ is to the future of $\lambda_1$. 
\item $(\lambda_1,\lambda_2)\in P^1_2$ if $\lambda_2 < P_{21}(\lambda_1)$; intuitively $\lambda_2$ is to the past of $\lambda_1$.
\item $(\lambda_1,\lambda_2)\in S_{12}$ if neither of these conditions is satisfied: intuitively $\lambda_2$ and $\lambda_1$ are not temporally ordered. 
\end{itemize}

\begin{figure}\label{fig_2wl_hs}
\def\svgwidth{0.4\columnwidth}
\begin{center}
\scriptsize
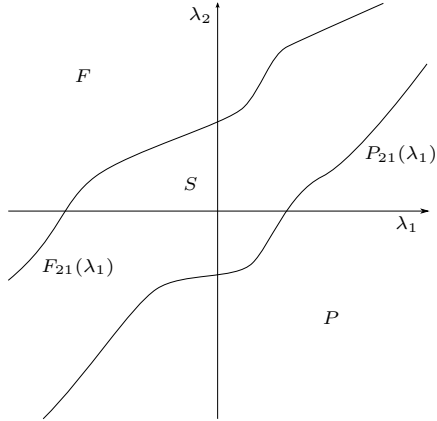
\normalsize
\end{center}
\caption{The two-worldline history space}
\end{figure}

\begin{figure}\label{fig_3_2wl_2IRs}
\def\svgwidth{0.9\columnwidth}
\begin{center}
\scriptsize
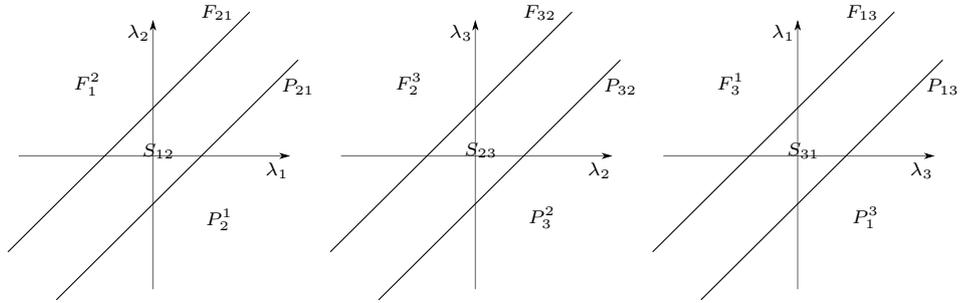
\normalsize
\end{center}
\caption{Six functions defining a three-worldline 2IR, and the resulting regions of history space}
\end{figure}

The 3-worldline history space $\mathcal{H}_3 = W_1\times W_2\times W_3$ consists of triples of points, one from each worldline. Just as with the 2-worldline case, the 2IR functions partition the history space into different regions; in this case there is considerably more complexity. It's instructive to approach this graphically. There are now six 2IR functions each of which defines a surface in the history space. Consider a very simple collection of functions as shown in figure \ref{fig_3_2wl_2IRs}. 
Each of these is extended `vertically' into the third dimension, to make a `wall'. Thus we have three walls all intersecting at angles. Each wall partitions the history space into three regions, as seen above, and the intersections of these regions generate multiple new regions. 

\begin{figure}\label{fig_3wl_hs}
\small
\begin{center}
\def\svgwidth{0.6\columnwidth}
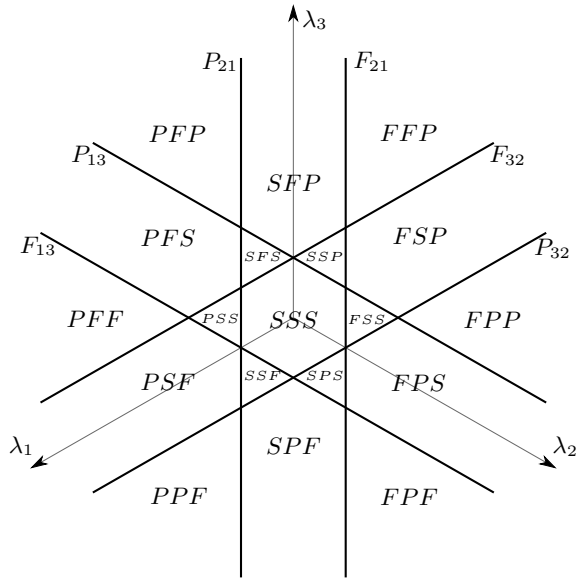
\end{center}
\normalsize
\caption{View onto a 3-worldline history space, with labelled regions}
\end{figure}

So what is the structure of the history space? Figure \ref{fig_3wl_hs} shows one view onto $\mathcal{H}_3$, looking towards the origin along the line $\lambda_1 = \lambda_2 = \lambda_3$. For readability we have dropped the super- and subscripts on the $F$s, $P$s and $S$s which label the regions\footnote{This could lead to some ambiguity: we need to remember that e.g. $FFP$ means $\lambda_2$ is to the future of $\lambda_1$, $\lambda_3$ is to the future of $\lambda_2$ and $\lambda_1$ is to the past of $\lambda_3$.}.
19 different regions appear in this diagram. Recall that the possible regions of $\mathcal{H}_3$ take the form $R_{12} R_{23} R_{31}$ where $R_{ij} \in \{P^i_j, S_{ij}, F^j_i\}$, so a priori, there are 27 possible regions. But in fact not all of these are consistent with the locality conditions defined in section \ref{sec_2IR_embed} which must be satisfied if a 2IR is to be embeddable. For example, suppose $(\lambda_1, \lambda_2, \lambda_3)\in FFS$. Then: (i) $F_{21}(\lambda_1) < \lambda_2$; (ii) $F_{32}(\lambda_2) < \lambda_3$; (iii) $\lambda_3 < F_{31}(\lambda_1)$. Taken together, we derive $F_{kj} \circ F_{ji} (\lambda_i)  < F_{ki}(\lambda_i)$, which violates the $F$ version of the locality condition, equation (\ref{eq_PP<P}). 

\begin{figure}\label{fig_deformed_star}
\small
\begin{center}
\def\svgwidth{0.9\columnwidth}
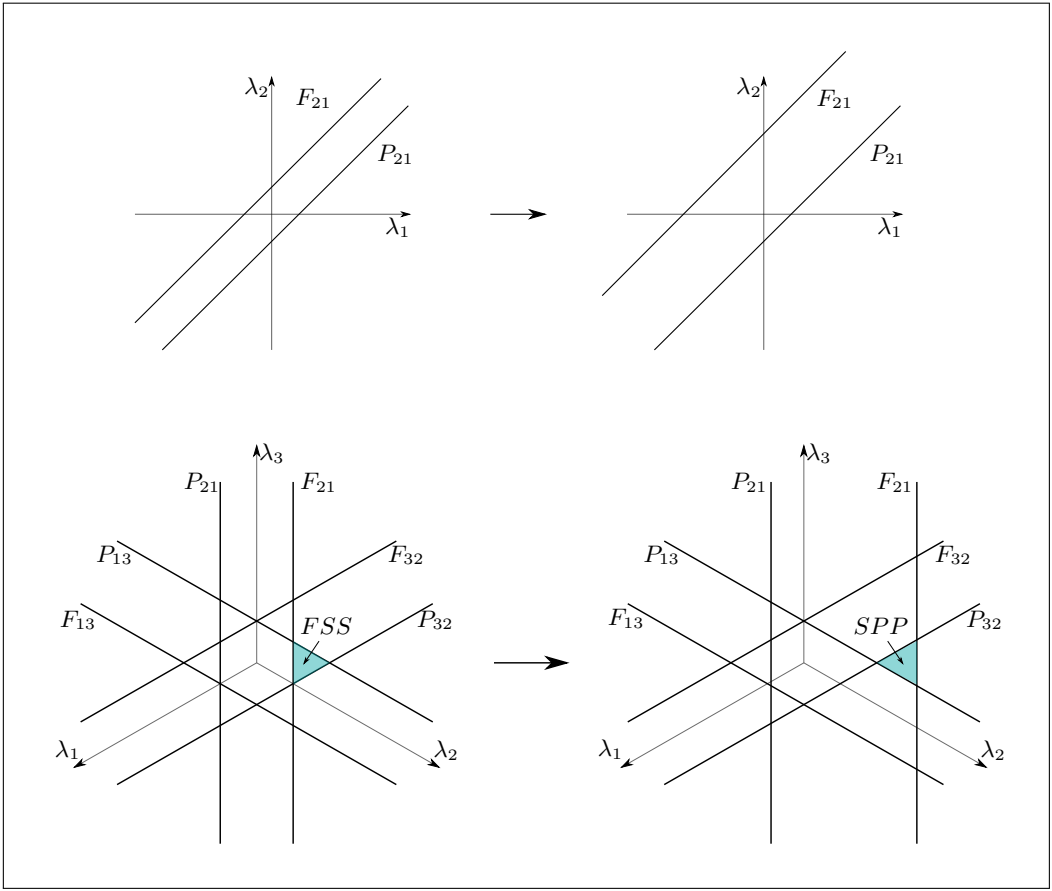
\end{center}
\normalsize
\caption{Altering one of the 2IR functions leads to non-embeddable regions appearing in $\mathcal{H}$}
\end{figure}

Similar arguments rule out the following 8 regions: 
\begin{itemize}
\item $(FFF)$
\item $(FFS)$, $(FSF)$, $(SFF)$
\item $(PPS)$, $(PSP)$, $(SPP)$
\item $(PPP)$
\end{itemize}
The rule is: \emph{if P appears twice the third character must be F, and if F appears twice the third character must be P}. 
The first and last regions imply the more drastic `Escher-like' failure of embeddibility denoted by failure of equation (\ref{eq_FF>P}). 
If the history space of a 3-worldline 2IR contains any of these 8 regions, then we cannot embed it fully into spacetime, although parts of it may still be embeddable. 
In figure \ref{fig_3wl_hs} there are 19 regions, and these are exactly the $19=27-8$ regions which are allowed by embeddibility. This is simply a consequence of the particular form of the six 2IR functions however. 
If, as shown in figure \ref{fig_deformed_star}, we slightly alter the $F_{21}$ function, a `forbidden' region ($SPP$) appears in the history space, and we lose an allowed region ($FSS$). 

\subsection{Projections onto history space}

\begin{figure}\label{fig_via_functions}
\small
\begin{center}
\def\svgwidth{0.4\columnwidth}
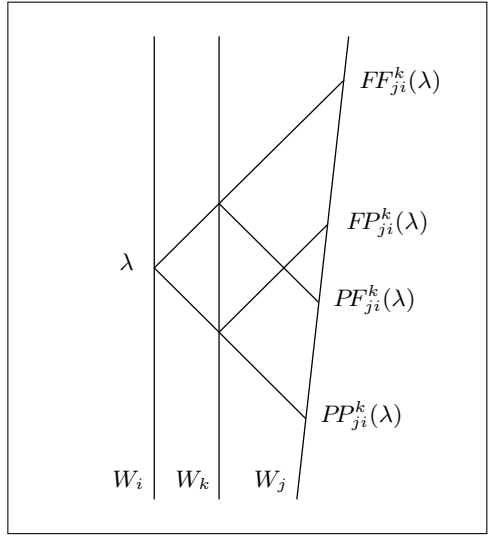
\end{center}
\normalsize
\caption{The via-$k$ functions}
\end{figure}

It's difficult to visualise the full three-dimensional history space, and we will mostly prefer to work with \emph{projections} onto the history space. The $(i,j)$-projection onto the 3-worldline history space is $W_i\times W_j$ partitioned into regions by a collection of six functions mapping $W_i$ into $W_j$: 
\begin{itemize}
\item The 2IR past and future functions $F_{ji}$ and $P_{ji}$; 
\item The \emph{via-$k$} functions: 
\begin{equation}F\!F^k_{ji} :=F_{jk}\circ F_{ki}\end{equation}
\begin{equation}F\!P^k_{ji} :=F_{jk}\circ P_{ki}\end{equation}
\begin{equation}P\!F^k_{ji} :=P_{jk}\circ F_{ki}\end{equation}
\begin{equation}P\!P^k_{ji} :=P_{jk}\circ P_{ki}\end{equation}
where $k\neq i,j$ is the third worldline. 
\end{itemize}
The via functions are so-called because they map points on $W_i$ to points on $W_j$ `going via' $W_k$ (as shown in figure \ref{fig_via_functions}). Their significance will become clear shortly. 

\begin{figure}\label{fig_projection}
\small
\begin{center}
\def\svgwidth{0.65\columnwidth}
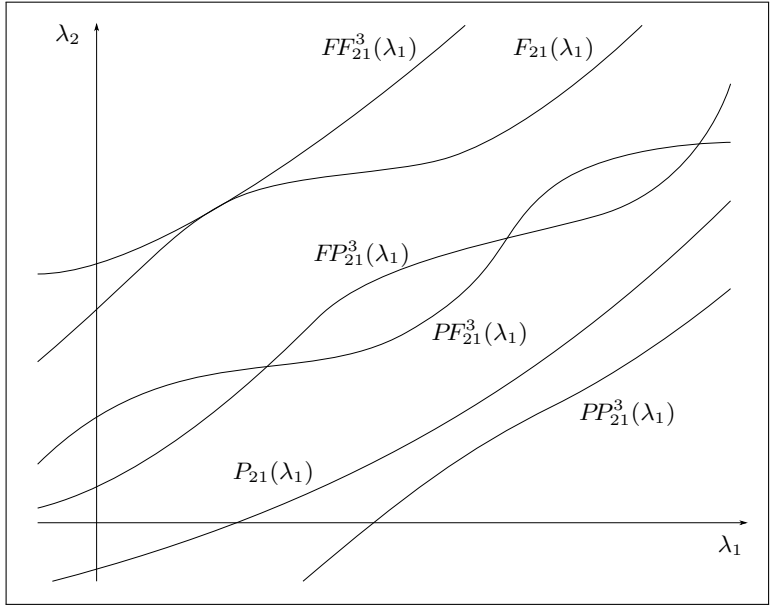
\end{center}
\normalsize
\caption{(1,2)-projection onto a 3-worldline history space}
\end{figure}

An example of a possible $(1,2)$-projection is given in figure \ref{fig_projection}. 
How generic are the orderings and intersections of the 2IR and via-3 functions in this figure? In the $(i,j)$-projection we have: 
\begin{itemize}
\item $F_{ji}>P_{ji}$, by definition. 
\item $F\!F^k_{ji}>F\!P^k_{ji}>P\!P^k_{ji}$ and $F\!F^k_{ji}>P\!F^k_{ji}>P\!P^k_{ji}$, (by the previous point, and by monotonicity of $F_{ij}$ and $P_{ij}$). 
\item There are no requirements on the ordering of $F\!P^k_{ji}$ and $P\!F^k_{ji}$. 
\item In general there are no restrictions on the ordering of the two 2IR functions w.r.t. the via functions. However, \emph{if the 2IR is embeddable} then we conclude from lemma \ref{lem_local_cond} that $F\!F^k_{ji}\geq F_{ji}\geq F\!P^k_{ji},P\!F^k_{ji}$, and $F\!P^k_{ji},P\!F^k_{ji}\geq P_{ji}\geq P\!P^k_{ji}$. 
\item If equality is satisfied in any of these conditions (i.e. if the curves touch or cross) then from lemma \ref{lem_collinearity} there is a collinearity in any embedding. Referring to the diagram in lemma \ref{lem_collinearity}, we note that if there is such an intersection in one projection, there will be one in each other projection.\footnote{For example if $F_{ji}(\lambda_i)=F\!F^k_{ji}(\lambda_i)$ in the $(i,j)$-projection, then $P_{kj}(\lambda_j)=F\!P^i_{kj}(\lambda_j)$ in the $(j,k)$-projection and $P_{ij}(\lambda_k)=F\!P^i_{ik}(\lambda_k)$ in the $(i,k)$-projection, where $\lambda_j=F_{ji}(\lambda_i)$ and $\lambda_k=F_{ki}(\lambda_i)$.}
\item If $P_{ji}\geq F\!F^k_{ji}$ or $F_{ji}\leq P\!P^k_{ji}$ then the 2IR fails to satisfy an Escher condition (definition \ref{def_Escher}) and exhibits a more extreme version of non-embeddibility. 
\end{itemize}
Returning to figure \ref{fig_projection}, we can see that this portion of the projection is consistent with an embeddable 2IR, and that we will have a collinearity in the embedding, corresponding to the point where the $F_{21}$ curve touches the $F\!F^3_{21}$ curve. 

An important issue is how much information the projections carry about the 2IR. In fact it is simple to see that one projection alone is insufficient to reconstruct the full 2IR, but given any two out of the three projections, we can do it. For example, given the $(1,2)$- and $(2,3)$-projections, we immediately have $F_{21}$, $P_{21}$, $F_{23}$ and $P_{23}$, and we can derive the remaining 2IR functions by e.g. $F_{31} = F\!F^1_{32}\circ P_{21}$, $P_{31} = P\!P^1_{32}\circ F_{21}$. This is important because it allows us to work exclusively with the easier-to-visualise projections. 
Note also that within one projection, given three of the via functions we can deduce the fourth. For example: 
\begin{equation}
P\!F^k_{ji}\circ (F\!F^k_{ji})^{-1} \circ F\!P^k_{ji}  =  (P_{jk}\circ F_{ki}\circ P_{ik}\circ P_{kj}\circ F_{jk}\circ P_{ki}) =  P\!P^k_{ji}
\end{equation}
This ties in with the previous point: each projection has five continuous monotonic degrees of freedom, and the 2IR has six; thus we need at least two projections to give us all the information about the 2IR. Explicitly, one can show that with the five independent functions of one projection, and just one function from another projection the entire 2IR can be reconstructed.\footnote{One possible set from which the whole 2IR can be recovered is $\{F_{21}, P_{21}, F\!F^3_{21}, F\!P^3_{21}, P\!F^3_{21}; P_{32}\}$. }

\subsection{Equivalence of 3-worldline 2IRs}

Recall the discussion in section \ref{sec_2IR_equivalence}: we believe that an $n$-worldline 2IR has $n(n-2)$ continuous monotonic degrees of freedom. In the case of three worldlines, we have three functional degrees of freedom: a 2IR has six future and past functions, but three of these can be put into any form we wish by reparameterisations of the worldlines. 

How do these three degrees of freedom look in terms of projections? Recall that for any 2IR, the five independent functions in the $(i,j)$-projection, and one function in the $(j,k)$-projection are enough to fix all the other projection functions. Now we consider 2IRs up to equivalence: by re-parameterising $W_i$ and $W_j$ we can fix any two of the functions in the $(i,j)$-projection, leaving three degrees of freedom. We can then fix the remaining function in the $(j,k)$-projection by re-parameterisation of $W_k$ (perhaps surprisingly, re-parameterisations of $W_k$ have no effect on the form of the via-$k$ functions in the $(i,j)$-projection). Thus the three functional degrees of freedom characterising an equivalence class of 2IRs can all be exhibited in a single projection.

\section{The Machian $\mathcal{R}\rightarrow\mathcal{I}$ construction}

Having explored a particular choice of relational configuration, we now turn back to more general considerations, which will nevertheless be important once we embark on trying to apply dynamics to 2IRs. 
We described in section \ref{sec_rel_ind} how relational configurations can sometimes be derived from individualist configurations, in effect, giving a function $D:\mathcal{I}\rightarrow\mathcal{R}$ from the individualist configuration space to the relational configuration space. 
Our hypothesis is that the world is fundamentally described by relational configurations, but that for certain collections of systems an individualist description is an accurate approximation to the real physics. To implement this idea we need to demonstrate a construction `converse' to the one described above, this time deriving individualist configurations from relational configurations. 

Given a relational configuration space $\mathcal{R}$, one might be tempted to look for $\mathcal{I}$ and $D:\mathcal{I}\rightarrow\mathcal{R}$, and then invert $D$. Whilst $\mathcal{I}$ and $D$ can often be found, typically $D$ is many-to-one and thus not invertible. For example assume that $\mathcal{R}$ is a $r_{ij}$-configuration space. Now take $\mathcal{I}$ to be a $\mathbf{x}_i$-configuration space, assigning a 3-dimensional vector to each particle. Then we have a map $D:\mathcal{I}\rightarrow\mathcal{R}$ defined via $r_{ij} = |\mathbf{x}_i - \mathbf{x}_j|$. However, note that any two individualist configurations $I,I' \in\mathcal{I}$ related by rigid translations and rotations in the background space have the same inter-particle distances, and thus $D(I) = D(I')$. 
We could make an arbitrary choice of configuration $I_R$ from each $D^{-1}(R)\subset\mathcal{I}$ to get $E:\mathcal{R}\rightarrow\mathcal{I}$ so that $E(R) = I_R$. But then the degrees of freedom of the individualist configuration $E(R)$ will not be fully determined by those of the relational configuration $R$. This will be a particular problem once we try to infer the fundamental relational dynamics from the individualist dynamics that we already know, because the individualist dynamics may employ individualist degrees of freedom that aren't fully determined by the underlying relational degrees of freedom via $E:\mathcal{R}\rightarrow\mathcal{I}$. An obvious example is when the individualist dynamics involves a default motion like inertial motion. 
We need a more direct method of constructing some $\mathcal{I}$ directly from $\mathcal{R}$. 

The problems that we noted above with dynamics suggest that we should base our construction on Mach's principle: given $I = E(R)$, the individualist degrees of freedom in $I_i$ should encode the relational degrees of freedom in $R$ pertaining to the $i^\text{th}$ body and a `background' of reference bodies. Furthermore, the sort of individualist configurations we want to derive are those which can be interpreted as locating all the bodies in a common space i.e. those for which all $I_i$ are isomorphic to some $I_0$, and we would like the construction to supply an account of what meaning `space' could have if relational configurations are fundamental. A relationalist claims that all that exists are material bodies and their spatial relationships, and that space has no fundamental existence. Yet he must still concede that the notion of an empty point in space is a coherent and useful idea. But if space does not fundamentally exist, then what are these points? Note that a particle placed at a point $p$ in space has a particular set of distances to all other bodies, and this set of distances is unique to the point $p$. So perhaps `space' as we familiarly understand it can be seen as the space of all sets of distances that a single body can have with all other bodies. There are as many of these distances as there are bodies, so in principle this space of `distance sets' should have very large dimension. But the relational configurations that we see are very tightly constrained, so that the space of distance sets is in fact 3-dimensional. 

We aim to formalise this intuition, and integrate key ideas from Mach's principle. We present only a framework for a type of $\mathcal{R}\rightarrow\mathcal{I}$ construction which we term Machian. For any given RCS the details will likely be quite specific. Note also that we are assuming that the relational degrees of freedom of $\mathcal{R}$ pertain to \emph{pairs} of systems, not larger $n$-tuples. It may be possible to loosen this restriction. The construction proceeds in several steps: 

\begin{enumerate}

\item
We must be able to identify a subspace $\mathcal{R}_0\subset\mathcal{R}$ such that: 
(1) $\mathcal{R}_0$ is symmetric under permutations of systems i.e. if $R\in\mathcal{R}_0$ then so is any configuration $R'$ obtained from $R$ by permuting systems;  
(2) The dimensionality is such if we choose $m$ arbitrary systems as reference systems then it remains to specify the relationship of each other system with these reference systems to completely identify the configuration. 
The point of this second requirement is that for some (non-reference) system $x$ the $m$ degrees of freedom pertaining to its relationships with the reference systems can be seen as properties of $x$, from which all its other relationships can be determined. 

\begin{itemize}
\item 
\textbf{Step 1:} The construction is only defined on $\mathcal{R}_0$. 
\end{itemize}

\begin{example}
As an example, let $\mathcal{R}$ be an $r_{ij}$-configuration space of 4 particles. $\mathcal{R}$ is 6-dimensional (since there are six interparticle distances). Requiring that the $r_{ij}$s can be interpreted as distances in 2-dimensional Euclidean space is equivalent to enforcing the vanishing of the Cayley-Menger determinant \cite{Blumenthal}: 
\begin{equation} 
\left| \begin{array}{ccccc}
0 & 1 & 1 & 1 & 1 \\
1 & 0 & r_{12}^2 & r_{13}^2 & r_{14}^2 \\
1 & r_{12}^2 & 0 & r_{23}^2 & r_{24}^2 \\
1 & r_{13}^2 & r_{23}^2 & 0 & r_{34}^2 \\
1 & r_{14}^2 & r_{24}^2 & r_{34}^2 & 0 \\
\end{array}\right| = 0
\end{equation}
which restricts us to a 5-dimensional subspace $\mathcal{R}_0$ of $\mathcal{R}$. 
Suppose we specify that $R\in\mathcal{R}_0$. Choose particles 1 and 2 as reference systems, and fix $r_{12}$. Now it suffices to give the relationships with the reference systems of particle 3 ($r_{13}$ and $r_{23}$) and particle 4 ($r_{14}$ and $r_{24}$) to determine all other relationships (in this case just the distance $r_{34}$).
\end{example}

\begin{example}\label{ex_5into2}
Actually the final sentence of the previous example is not quite correct: in fact $r_{34}$ is determined to be one of two discrete values. To see this, note that once particles 1 and 2 are fixed in 2D space, specifying $r_{13}$ and $r_{23}$ determines the position of particle 3 up to reflection in the line connecting particles 1 and 2. The same goes for particle 4, and thus there is still uncertainty as to whether particles 3 and 4 are on the same or different sides of this line. Thus in this case although there are two degrees of freedom associated with each particle, we actually need to give their distances to \emph{three} reference systems.
Therefore we now consider a configuration of five particles. $\mathcal{R}$ is 10-dimensional, and constraining to embedding in 2-dimensional space requires several Cayley-Menger determinants to vanish, ultimately resulting in a 7-dimensional $\mathcal{R}_0$. 
\end{example}

\textbf{Note:} If this construction is to be useful in the description of real physics we will have to at some point justify why we should be restricted to configurations in $\mathcal{R}_0$, at least in those cases when we recover classical theories. One might hope that it could be derived on dynamical grounds, perhaps with the constraint surface forming an attractor in the phase space. 

\item
It must be possible to partition the $N$ systems into (i) a collection of $n$ systems whose motion is of interest, the \emph{actors} and (ii) a collection of $N-n$ systems whose relationships with each other are essentially static over some suitable time-scale, which constitute the \emph{background}. We must have $m \leq N-n$ so that all the reference systems referred to above can be made part of the background. 

\begin{itemize}
\item
\textbf{Step 2:} We restrict to a subspace $F\mathcal{R}_0$ whose configurations all have the same set of $b$-$b$ relationships - we \emph{freeze the background}. 
\end{itemize}

The construction will work with any choice of $b$-$b$ values, but the form of any induced dynamics on the individualist configurations that result from the construction will likely depend heavily on the choice (as we might expect from Mach's principle). 
$F\mathcal{R}_0$ is the space of configurations which can actually arise. But these configurations can be repackaged into an ICS as we now describe. 

\begin{example}\label{ex_freeze}
In example \ref{ex_5into2} $\mathcal{R}_0$ was 7-dimensional. We will choose particles 1, 2 and 3 to be our background, freezing the values of $r_{12}$, $r_{23}$ and $r_{13}$ to get a 4-dimensional $F\mathcal{R}_0$. 
\end{example}

\textbf{Note:} Of course once again we must be able to argue why the partitioning into actors and background is physically reasonable. Again, justification must ultimately come from the dynamics, but it seems reasonable that the evolution of some systems (the background) might not be strongly influenced by other systems (the actors) - perhaps the background bodies have a much larger inertia. 
Note that $\mathcal{R}$ has three types of degrees of freedom which we will denote as $a$-$a$, $a$-$b$, and $b$-$b$, depending on whether the two systems are both actors, one is an actor and one is background, and so on. 

\item
In the final step we `quotient out' all degrees of freedom from $F\mathcal{R}_0$ which do not pertain to the $i^\text{th}$ actor. 

\begin{itemize}
\item
\textbf{Step 3:}
Consider the $i^\text{th}$ actor system, $a_i$. There is an equivalence relation $\sim_{a_i}$ on $F\mathcal{R}_0$ which relates configurations which agree on all $a_i$-$b$ degrees of freedom. The space of these equivalence classes $F\mathcal{R}_0/\sim_{a_i}$ is $\mathcal{I}_{a_i}$. 
\end{itemize}

Each point in $\mathcal{I}_{a_i}$ is a different collection of $a_i$-$b_j$ degrees of freedom from $F\mathcal{R}_0$. 
Given $\mathcal{I} = \mathcal{I}_{a_1} \times\dots\times \mathcal{I}_{a_n}$ we can reconstruct $F\mathcal{R}_0$, because: (i) we earlier required of $\mathcal{R}_0$ that all the relationships pertaining to a non-reference system could be deduced from its relationships with the reference systems; and (ii) we required that there are at least enough background systems to act as the reference systems. Thus $\mathcal{I}$ and $F\mathcal{R}_0$ are equivalent. 

Furthermore, because of the symmetry under permutations of $\mathcal{R}_0$, exactly the same combinations of values for the $a_i$-$b_j$ degrees of freedom occur in each $\mathcal{I}_{a_i}$: thus we have a natural bijection between all the $\mathcal{I}_{a_i}$s, indicating that they describe the same `space' $\mathcal{I}_0$. 

\begin{example}
Continuing examples \ref{ex_5into2} and \ref{ex_freeze}, to get $\mathcal{I}_4$ we identify all points of $F\mathcal{R}_0$ which share the same combination of values for $r_{14}$, $r_{24}$, and $r_{34}$. As explained in example \ref{ex_5into2}, despite needing all three of these distances to parameterise it, this space is 2-dimensional.  
$\mathcal{I}_5$ is formed similarly and $\mathcal{I}=\mathcal{I}_4\times\mathcal{I}_5$. 
\end{example}

\end{enumerate}

So what the construction actually does is this: we begin with a configuration $R\in\mathcal{R}_0$ of all $N$ systems (actors and background), and derive an individualist configuration $I$ of just the $n$ actors, interpretable as the actors all occupying points of the same `space' $\mathcal{I}_0$. 
This background space $\mathcal{I}_0$ is only a meaningful notion when applied to the actors, but this does not preclude it from playing the role of `space' or `spacetime' as found in our familiar classical theories since as Mach observed, these are only well-tested for small subsystems of the universe. 

The construction is designed to match the way that space actually appears to us: multiple moving bodies (including ourselves) constantly changing their relationships with a background of approximately static bodies (be they the trees and buildings around us, or the distant stars). 

The $\mathcal{R}\rightarrow\mathcal{I}$ construction for 2IRs remains to be worked out in any detail. However, we can draw some rough conclusions. There are four dimensions of spacetime. This suggests that if $W_i$ is the worldline of an actor, and $\lambda_i$ is a point on that worldline, we need only specify four of the $2(N-1)$ worldline instants $f_{ji}(\lambda_i)$ and $p_{ji}(\lambda_i)$ to determine all of them. The points of 4-dimensional spacetime would then essentially be sets of instants on other worldlines to which an instant on your worldline could be connected, via its $f_{ji}$ and $p_{ji}$ functions. A detailed analysis awaits a fuller understanding of the constraints that determine whether or not a 2IR embeds into Minkowski spacetime.

\subsection{Relevance to dynamics}\label{sec_R->I_dyn}

We are attempting to guess explicitly relational configurations and their dynamics such that under some circumstances we recover familiar classical theories, phrased in terms of individualist configurations. The Machian $\mathcal{R}\rightarrow\mathcal{I}$ construction now suggests the following approach to guessing relational dynamics, which manifestly implements Mach's principle. As we have mentioned, many theories employing individualist configurations have dynamics with two elements: a default evolution on each $\mathcal{I}_i$, and extra interaction terms typically expressed in terms of derived relationships between systems. 
For example in classical physics we typically have equations of motion with two types of term, inertial and force/interaction: 
\begin{equation}
\frac{\text{d}p_i}{\text{d}t} = \sum_{j\neq i} F_{ij}
\end{equation}
But if $\mathcal{I}$ is constructed using the Machian procedure then the degrees of freedom of $\mathcal{I}_i$ represent a summary of the $i^\text{th}$ system's relationships with the background systems. 
The Machian construction suggests that the terms on both sides of this equation should derive from fundamentally similar terms in the relational theory, the LHS representing $a$-$b$ interactions, the RHS representing $a$-$a$ interactions, with the apparent difference between the terms resulting from the differing treatment of actors and background in the $\mathcal{R}\rightarrow\mathcal{I}$ construction.

\section{Summary and further work}

We have introduced the idea of relational and individualist configurations, and argued that non-embeddable relational configurations could be relevant for physics in explaining why notions of space and time seem to break down in theories such as quantum mechanics and general relativity. The aim would be to define explicitly relational configurations which are typically non-embeddable, and a dynamics for these configurations, such that the configurations of certain collections of systems are forced to be embeddable, and we recover the familiar individualist theories of classical physics for these systems. 
We presented a type of explicitly relational configuration, the 2IR, from which the familiar individualist configurations of timelike worldlines in spacetime can potentially be recovered. We investigated the properties of 2IRs, particularly with regard to their embeddibility into spacetime. 
Finally we considered the more general issue of how individualist configurations can be derived from relational configurations in a manner that introduces no arbitrary data and which manifestly implements Mach's principle, and we speculated on its implications for relational dynamics. 

These results are very preliminary, and a great deal remains to be done to make this a viable programme of research. There are several pressing issues. 
\begin{itemize}

\item \textbf{Embedding 2IRs into Minkowski space:}
We would like to derive constraints on 2IRs which determine whether they embed into Minkowski spacetime. For those which do embed, we want to know how uniquely they embed, or put another way, how much of the individualist spacetime configuration we can infer from the 2IR. As we saw in section \ref{sec_2IR_embed} we may be able to make progress on both counts by calculating the intersections of ellipsoids in Minkowski space. Also useful might be the explicit calculation of the 2IRs deriving from various configurations of worldlines in spacetime. We also have not yet considered the question of embedding for $\tau$-2IRs. 

\item \textbf{Dynamics for 2IRs:}
Dynamics is often taken to mean evolving configurations in time, but in a `relativistic' setting such as the 2IR where the configuration is a complete history, the role of dynamics is more to single out some configurations as physically realisable. We could attempt to define dynamics from scratch, but perhaps more promising is to look at the dynamics of theories we are hoping to derive and `reverse-engineer' the dynamics for the fundamental relational configurations. The dynamics of the WF-EM theory are shown in \cite{WF49} to be equivalent to the conservation along each worldline of a quantity of the form: 
\begin{equation}
\sum_{i} G^\mu_i(\lambda_i) + \sum_{i<j} G^\mu_{ij}(\lambda_i,f_{ji}(\lambda_i)) +  \sum_{i<j} G^\mu_{ij}(\lambda_i,p_{ji}(\lambda_i)) 
\end{equation}
where $G^\mu_i(\lambda_i)=m_i c \dot{x}^\mu(\lambda_i)$ is equal to the 4-momentum of worldline $W_i$ at the instant $\lambda_i$, and the second and third terms are associated with pairs of instants on $W_i$ and $W_j$ connected by null geodesics, i.e. pairs of instants connected by the $f_{ji}$ and $p_{ji}$ functions of the derived 2IR (for full details see \cite{WF49} p.430). Recalling the arguments of section \ref{sec_R->I_dyn}, we propose that the single particle 4-momenta terms in fact derive from $G^\mu_{ik}$-type terms, where $W_k$ is a background worldline, via a Machian $\mathcal{R}\rightarrow\mathcal{I}$ construction. In rough terms, in 2IR dynamics energy and momenta would no longer be associated with individual worldlines, but with the $f_{ji}$ and $p_{ji}$ functions that connect pairs of worldlines. The details remain to be worked out. 

\item \textbf{The Machian $\mathcal{R}\rightarrow\mathcal{I}_0$ construction for 2IRs:}
Given the preceding discussion of dynamics, it is clearly essential to work out the details of the $\mathcal{R}\rightarrow\mathcal{I}_0$ construction which derives Minkowski spacetime from an appropriate 2IR. A key question is what form the relationships amongst the background worldlines must take. The background must be quite `uniform' in some sense, if we are to recover a default motion which matches inertial motion in Minkowski space. 

\item \textbf{Origin of embedding constraints:}
The imposition of embedding constraints on relational configurations is essential for reformulating them in individualist terms. But there is no a priori reason for these constraints to hold and for us to be restricted to embeddable configurations. It is incumbent upon a relationalist to provide an explanation for why the configurations we commonly encounter are embeddable, if he believes that they are fundamentally relational - failure to do so would suggest that the configurations are fundamentally individualist. Thus, if we begin with explicitly relational configurations we must provide an explanation for why the embedding constraints hold for the objects of classical physics. At present we have no detailed proposals for this. The best hope seems to lie in the as yet unformulated dynamics: perhaps for certain classes of systems the subspace of embeddable configurations is an attractor for dynamical trajectories. 

\item \textbf{Embedding into more general spacetimes:}
Ultimately we want to recover general relativity, and thus we will have to consider the issue of embedding our relational structures into more general spacetimes. The key question that arises is whether we can infer details of the geometry of the spacetime from the 2IR? If so, how much? For example, if the metric tensor could be inferred from the 2IR, then dynamical equations for the 2IR functions might potentially yield field equations for the metric. 

First note that a 2IR embeds into a spacetime as a set of timelike worldlines and the null geodesics connecting them. Thus the information contained within it is at best a subset of the causal structure of the spacetime\footnote{The causal structure of a spacetime is a relation on the points of the spacetime, corresponding to whether or not a pair of points can be connected by a causal curve (i.e. a curve whose tangent vector is nowhere spacelike). See \cite{H&E} chapter 6 for details.}. For most physically reasonable spacetimes, the causal structure determines the metric up to conformal equivalence\footnote{Two metric tensors $g_{\mu\nu}$ and $g'_{\mu\nu}$ are conformally equivalent if $g'_{\mu\nu}(x) = \Omega^2(x)g_{\mu\nu}(x)$; $\Omega^2(x)$ is termed the \emph{conformal factor}.}\cite{HKM, Malament}. Thus it seems that we would at best be able to recover the metric up to conformal factor from the 2IR. One could actually argue that this is what we would expect from Mach's principle. Conformally equivalent spacetimes have the same causal structure, but do not agree on which timelike curves are geodesics; in general relativity, timelike geodesics are the worldlines of inertially moving particles; and according to Mach's principle inertial motion should only make sense for small subsystems of `actors' in a large universe. This suggests that perhaps we should not expect to be able to deduce the conformal factor of the spacetime into which we embed the complete 2IR. 

To what extent could the causal structure of the spacetime be recovered from the 2IR? Clearly one could only hope to recover it up to a certain `resolution': we can't expect to determine the structure of the spacetime on scales much smaller than the separation of the closest worldlines. 
A first step is to try deriving 2IRs from configurations of worldlines in spacetimes that are not conformally flat. It swiftly becomes apparent that the 2IR in its current form may not be the correct structure to be working with in this case. It was defined by analogy with light-cones in Minkowski spacetime, which intersect any given timelike worldline at a single instant along its length. But in a general spacetime the light-cone of one instant on a worldline might intersect another timelike worldline at multiple points - consider for example the circular `photon orbits' of the Schwarzschild solution. We may need a generalised 2IR with more than the two functions $f_{ij}$ and $p_{ij}$ defined between each pair of worldlines. 

Finally we should note that the problem of finding a spacetime into which a structure can embed consistent with the causal structure is one of the key issues faced by the \emph{causal sets} approach to quantum gravity \cite{Causets}, and results in this field may be of use for our problem. 

\item \textbf{Connection to quantum theory:}
We began this paper with some motivational examples which referenced quantum theory, but have otherwise made little mention of it. One of the original motivations for this line of work was the hope that non-embeddable relational configurations could be the basis for a realist underpinning for quantum mechanics (a non-local `hidden variable' theory). 
The essential idea is that quantum mechanics would be the description of the interaction of classical systems -- whose relationships would embed into a background, and for whom classical concepts such as position and momentum would be well-defined -- with microscopic systems -- whose relationships would not be accommodated by this background, and therefore for whom classical concepts such as position would not be well-defined at most times. From this point of view the embeddable configuration of classical systems is as important to deriving quantum mechanics as is the microscopic `quantum' system, and progress on this idea really waits on a better understanding of 2IR dynamics and how this could enforce embeddibility for some but not all systems. 
Even if we are nowhere near a derivation of quantum theory, we might be able to make progress replicating some of the phenomenology of quantum mechanics. 
Recall that the `locality' conditions (equations \ref{eq_PP<P} to \ref{eq_PF>P}) of section \ref{sec_2IR_embed} need to be satisfied for all triples of worldlines in order for a 2IR to embed into Minkowski spacetime. A violation of these conditions for some triples of worldlines in an otherwise embeddable 2IR would imply that the instants paired by the 2IR on certain worldlines would be spacelike separated in any embedding. This could potentially be used  to explain the failure of local causality evidenced by violation of Bell's inequalities (see e.g. \cite{BellNC}), although we would still need to explain why these superluminal connections could not be exploited for signalling. 

\end{itemize}

We conclude with a brief discussion of connections between this and other work. A significant inspiration for this work came from Barbour and Bertotti's early work on relational particle dynamics, principally the BB77 theory \cite{BB77} which has been mentioned several times herein. Barbour has continued to work on relational and `Machian' theories with many collaborators. However, beginning with his second paper with Bertotti \cite{BB82} he has worked principally with what we have termed ID-relational configurations, which do not allow for non-embeddibility. Subsequent work by Barbour and others has extended this programme to general relativity; this in turn inspired the theory of \emph{shape dynamics} \cite{SD} originally due to Koslowski, Gomes and Gryb; in all of this work the fundamental objects are fields, not particles, and it has a `continuous' character which is lacking in the 2IR and other non-embeddable configurations. 
Mention has already been made of the \emph{causal sets} programme \cite{Causets}, championed by Sorkin and many others, which begins with a discrete structure (a locally finite partially ordered set) and tries to recover continuous spacetime as an approximate description of this structure, with a key issue being to determine whether the discrete structure can be embedded in any given spacetime. Despite this formal similarity with our approach, the causal sets programme is fundamentally different in its philosophy and aims. The elements of the causal set are taken to be discrete precursors to spacetime itself, so the theory is essentially substantivalist. Furthermore, their ultimate aim is to define a `sum over histories' of causal sets, thus `quantising spacetime'. In contrast, our hope is ultimately to show that quantum mechanics arises as a description of the interaction of embeddable classical measuring devices with non-embeddable microscopic system, and that quantum ideas should not play a role in fundamental theories. 
Markopoulou and Smolin \cite{QTfromQG} take a somewhat similar view to us on quantum mechanics. Here the fundamental objects are the vertices of a graph, with locality at a fundamental level corresponding to adjacency on the graph. Under low energy conditions the graph may embed into a low-dimensional space, but there is no requirement that `nearness' in this space correlates with adjacency in the underlying graph. Thus non-locality arises as an artefact of a non-faithful embedding of a discrete relational structure into a continuum. 
The work of Amelino-Camelia, Friedel, Kowalski-Glikman and Smolin on \emph{relative locality} \cite{RelLoc} has certain similarities with our approach. The fundamental objects are particles, and spacetime is an emergent rather than fundamental notion. One peculiarity of this approach is that each particle sees a `different spacetime'. Something rather similar might happen in the 2IR case with different partitions of the worldlines into actors and background. Whilst spacetime is emergent in this approach, energy and momentum are fundamental attributes of particles, and are crucial to the dynamics, consideration of which may aid in the development of 2IR dynamics. 
Finally we note that Westman and Sonego in \cite{Hans} develop a view of spacetime that is very reminiscent of our Machian $\mathcal{R}\rightarrow\mathcal{I}$ construction: they consider all scalars derivable from the metric and matter fields of GR, then form the space of all combinations of values which these scalars can take; it then turns out that for any given model the allowed combinations of values form a 4-dimensional subspace of this space, which can be interpreted as spacetime.

\section*{Acknowledgements}

I am grateful to Lucien Hardy for many useful conversations and for his support with this project. 
I have also profited from discussions with Julian Barbour, Rob Spekkens, Rafael Sorkin, Laurent Friedel and Lee Smolin. 
Research at Perimeter Institute is supported in part by the Government of Canada through NSERC and by the Province of Ontario through MRI.


\bibliographystyle{plain}
\bibliography{nerc}

\end{document}